\documentclass[11pt,a4paper,oneside]{article}
\usepackage[top=3cm, bottom=3cm, left=2cm, right=2cm]{geometry}
\usepackage[english]{babel}
\usepackage[utf8]{inputenc}
\usepackage[T1]{fontenc}

\usepackage{amsthm,amsmath,amssymb,mathrsfs,dsfont,mathtools}
\usepackage{bbm}
\usepackage{multirow}
\usepackage{bm}
\usepackage{graphicx}
\usepackage[colorlinks=true, allcolors=blue]{hyperref}
\usepackage{comment}
\usepackage[all]{nowidow}
\usepackage{booktabs}
\usepackage{microtype}  % adjust layout imperfections
\usepackage{xcolor}

\usepackage{algorithm}
\usepackage{algpseudocode}

\usepackage{etoolbox}
\AtBeginEnvironment{table}{\linespread{1}\selectfont}

\usepackage[authoryear, round]{natbib}
\usepackage{authblk}
\date{}
\title{Martingale posteriors for flexible regression via recursive least squares}
\author[1]{Davide Agnoletto}
\author[2]{Tommaso Rigon}
\author[1]{David B. Dunson}
\affil[1]{Department of Statistical Science, Duke University, Durham, NC, USA}
\affil[2]{Department of Economics, Management, and Statistics, University of Milano--Bicocca, 20126 Milano, Italy}

\providecommand{\keywords}[1]{
  \small
  \textbf{\textit{Keywords:}} #1
  \normalsize
}

\usepackage[sf]{titlesec}
\usepackage{sectsty}
\allsectionsfont{\centering \normalfont\scshape} % Make all sections centered, the default font and small caps

\newtheorem{theorem}{Theorem}

\newtheorem{lemma}{Lemma}
\newtheorem{proposition}{Proposition}
\theoremstyle{definition}
\newtheorem{assumption}{Assumption}

\newtheorem{remark}{Remark}
\theoremstyle{remark}
\newtheorem{example}{Example}

\usepackage{apptools}
\AtAppendix{\counterwithin{theorem}{section}}
\AtAppendix{\counterwithin{lemma}{section}}
\AtAppendix{\counterwithin{proposition}{section}}
\AtAppendix{\counterwithin{corollary}{section}}

\newcommand{\var}{\mathrm{var}}

\begin{document}

\maketitle

\begin{abstract}
Predictive Bayes quantifies uncertainty by specifying a sequence of one-step-ahead predictive distributions, inducing a martingale posterior without an explicit prior. Existing parametric constructions rely on stochastic-gradient updates and have seen limited development for flexible regression. We propose a predictive Bayes framework based on recursive least squares, whose martingale posterior has an explicit representation: given future covariates, it is the law of the posterior mean plus a weighted series of independent draws from the standardized predictive distribution. The posterior mean is a penalized least squares estimator, available in closed form and invariant to data ordering, whereas the predictive distribution, for example heavy tailed, is chosen separately without changing the estimator or algorithm. We derive the exact posterior variance under a known design and an asymptotically Gaussian approximation, and obtain the martingale posterior in closed form for Gaussian predictive distributions. Posterior draws are independent by construction, with a known truncation error, and require no Markov chain Monte Carlo. The construction extends to additive, mixed and Gaussian process models. In simulations and an analysis of \textsc{nhanes} triglyceride data, the method attains accuracy comparable to that of a fully Bayesian model with the same response distribution at a fraction of its computational cost.
\end{abstract}

\section{Introduction}

Flexible regression is central to modern statistical practice, yet Bayesian analyses in this setting can be awkward when the scientifically appropriate sampling distribution is not paired with a convenient prior or computational scheme. Heavy tails and rich mean structures are readily motivated in applications, but can complicate posterior computation. 
Laplace approximations \citep{rue2009inla} and variational Bayes \citep{blei2017vi} avoid Markov chain Monte Carlo (\textsc{mcmc}) in such settings, but their accuracy depends on the approximating family, and variational methods tend to understate posterior uncertainty. Generalized Bayesian approaches, which replace the likelihood with moment conditions or a loss \citep{chernozhukov2003mcmc, bissiri2016general}, typically rely on \textsc{mcmc}, and valid uncertainty quantification hinges on calibrating a weighting matrix or learning rate \citep{bissiri2016general, agnoletto2025bayesian}.
Predictive Bayes offers a different route to uncertainty quantification: rather than starting from a prior, it specifies a sequence of one-step-ahead predictive distributions. When the corresponding parameter estimates form a martingale, their random limit has an induced \emph{martingale posterior} distribution. This construction retains a Bayesian interpretation while making the sequential predictive rule, rather than an explicit prior, the primary modeling object \citep{fortini_petrone_2020, holmes2023statistical, fong_holmes_2023}.

The regression setting raises two obstacles for existing predictive Bayes methods. First, a useful regression procedure should allow the predictive distribution to reflect the response distribution while retaining a stable and interpretable estimator of the conditional mean. Second, its finite-sample estimator should not depend on an arbitrary ordering of the observed data. Much recent work develops predictive constructions for density estimation or discrete data \citep{fortini_petrone_2020, agnoletto2025nonparametric}. Parametric martingale posteriors have mainly been built from stochastic-gradient or online-gradient updates \citep{walker2022new, holmes2023statistical, fong2024asymptotics, fortini2024exchangeability}. These updates are valuable general tools, but they require delicate learning-rate choices and typically yield order-dependent finite-sample estimates. They also do not directly provide a flexible regression framework with transparent posterior uncertainty.

We address these issues by constructing martingale posteriors for regression from recursive least squares (\textsc{rls}). The \textsc{rls} recursion is a classical and computationally inexpensive estimator, but here it is used as the update of a predictive Bayesian model. The resulting posterior mean is invariant to permutations of the observed data and coincides with a ridge-type regression estimator. Crucially, the plug-in predictive distribution can be chosen separately from this update: it may be Gaussian or heavy-tailed, for example, while preserving the same conditional mean update, so that changing the response distribution requires neither a new estimator nor a new sampler. Posterior uncertainty is obtained by forward predictive resampling, which produces independent draws, with no convergence diagnostics and a truncation error known in closed form. Hyperparameters, such as smoothing parameters, can be selected by maximizing an order-invariant leave-one-out prequential log-likelihood. In our simulations and application, the resulting posterior mean has an accuracy comparable to that of a fully Bayesian model with the same response distribution, at a fraction of its computational cost.

Our main contributions are as follows. We show that the \textsc{rls} coefficient sequence is a martingale and converges to a random limit, establishing a martingale posterior with an order-invariant finite-sample posterior mean and asymptotic exchangeability of the induced sequence. Conditional on a known covariate design, we derive the exact posterior variance and establish an asymptotically Gaussian approximation that uses only observed covariates. Under Gaussian predictive distributions, we obtain the exact martingale posterior in closed form, and we show that, with a suitable predictive variance, the construction coincides with the conjugate Bayesian linear model, for any covariate distribution. Finally, we show how the construction can be used for additive modeling and to accommodate correlated coefficients, as in mixed effects and Gaussian process regressions.

\section{Predictive Bayes and martingale posteriors}
\label{sec:background}

Predictive Bayes starts from a sequential specification rather than an explicit prior. Let $(Y_n)_{n\geq1}$ be a sequence of random variables and write
\begin{equation*}
    Y_1\sim P_0,
    \qquad
    Y_{n+1}\mid y_{1:n}\sim P_n,\quad n\geq1,
\end{equation*}
where $P_n$ is the one-step-ahead predictive distribution after observing $y_{1:n}$. By the Ionescu-Tulcea theorem, this sequence of predictive distributions defines a unique joint probability law $\mathds{P}$ for $(Y_n)_{n\geq1}$ \citep{berti2025probabilistic}.
By de Finetti's theorem, if $(Y_n)_{n\ge1}$ is exchangeable, then it admits a representation in terms of a statistical model $\Tilde{P}$ and a prior distribution $\Pi$ such that $Y_n\mid\Tilde{P}\overset{\text{iid}}{\sim}\Tilde{P}$, for any $n\ge1$, with $\Tilde{P}\sim\Pi$, and moreover the sequence of predictive distributions $(P_n)_{n\ge1}$ converges weakly to $\Tilde{P}$, $\mathds{P}$-a.s.
Except for well-known cases, constructing sequences of predictive distributions that ensure exchangeability is challenging in practice. A more tractable goal is asymptotic exchangeability: if $P_n$ converges weakly to a random distribution $\tilde P$, $\mathds{P}$-a.s., then  $(Y_{n+1},Y_{n+2},\ldots)\xrightarrow{d}(Z_1,Z_2,\ldots)$ as $n\to\infty$, where $(Z_1,Z_2,\ldots)$ is an exchangeable sequence with directing random probability measure $\Tilde{P}$, and $\xrightarrow{d}$ denotes convergence in distribution \citep{aldous1983exchangeability, fortini2012predictive, fortini2024exchangeability}. Thus, a convergent predictive rule induces a well-defined distribution for the random limit~$\tilde P$.

For a parametric family $\{P(\cdot\mid\theta):\theta\in\Theta\}$, a convenient construction specifies a sequence of estimates $\theta_n=\theta(y_{1:n})$ and uses the plug-in predictive distribution $P_n(\cdot)=P(\cdot\mid\theta_n)$. 
Specifically, a proper Bayesian model is built by sequentially alternating
\begin{equation}
    \theta_n = \theta(y_{1:n}) \qquad\text{and}\qquad
    Y_{n+1}\sim P_n(\cdot) = P(\cdot\mid\theta_n),
    \label{eqn:seq_general}
\end{equation}
for $n\ge1$. If $(\theta_n)_{n\ge1}$ is an $L_1$-bounded martingale, namely
\begin{equation*}
    \mathds{E}(\theta_{n+1}\mid y_{1:n}) = \theta_n,\quad n\ge1,
    \qquad\text{and}\qquad
    \sup_{n}\mathds{E}(|\theta_n|) < \infty,
\end{equation*}
then by Doob's martingale convergence theorem,
there exists a random $\tilde{\theta}\in\Theta$ such that $\theta_n\to\tilde{\theta}$ $\mathds{P}$-a.s. as $n\to\infty$ \citep{doob1953stochastic}. 
When $P(\cdot\mid\theta)$ is continuous in $\theta$, the predictive distributions converge to $P(\cdot\mid\tilde\theta)$ and the conditional law of $\tilde\theta$ given $y_{1:n}$ is called the \emph{martingale posterior} \citep{walker2022new, holmes2023statistical}. The posterior is generally not available analytically. It can, however, be sampled by predictive resampling: 
starting from $P_n$, repeatedly sample future observations and
update the plug-in predictive distributions as in \eqref{eqn:seq_general}, obtaining $\theta_N$ for $N\gg n$.
Independent forward paths yield independent approximate draws from the martingale posterior \citep{fortini_petrone_2020, fong_holmes_2023}.

Existing parametric methods rely on predictive updates that resemble stochastic gradient descent \citep{holmes2023statistical, fong2024asymptotics, fortini2024exchangeability, sun2026design}.
Because the resulting $\theta_n$ typically depends on the order of the observed data and requires careful tuning of the learning rate, most of these methods use the predictive construction only for predictive resampling and compute the initial $\theta_n$ separately through a different scheme.
In the next section, we instead use recursive least squares as the predictive update.
This yields a posterior mean that is invariant to the ordering of the data, giving an effective method for flexible regression.

\section{Predictive linear regression via recursive least squares}
\label{sec:rls}

\subsection{A predictive regression model}
\label{sec:model}

We now construct a martingale posterior for the regression coefficient $\bm\beta\in\mathds{R}^p$. Let $\bm X_n\in\mathds{R}^p$ and $Y_n\in\mathds{R}$ denote the covariates and response, with observed values $\bm x_n$ and $y_n$. We take $\bm X_n$ to be iid from an unknown distribution $P_X$ independent of past responses.
At the end of this section, we also introduce the case of fixed covariates, which is used in Sections~\ref{sec:variance} and~\ref{sec:asymptotics}. All proofs are given in Appendix~A.

\begin{assumption}
\label{as:predictive}
Conditional on the covariates, the predictive distribution belongs to the location-scale family
$\mathcal{P}_{\beta}=\{P(\cdot\mid\bm x,\bm\beta,\sigma):\bm x\in\mathds{R}^p,\ \bm\beta\in\mathds{R}^p,\ \sigma>0\}$,
where $P(\cdot\mid\bm x,\bm\beta,\sigma)$ is the law of $\bm x^\top\bm\beta+\sigma\xi$ and $\xi\sim P_\xi$, a fixed distribution on $\mathds{R}$ with $\mathds{E}(\xi)=0$ and $\mathds{E}(\xi^2)=1$.
\end{assumption}
Hence, $P(\cdot\mid\bm x,\bm\beta,\sigma)$ has mean $\bm x^\top\bm\beta$ and variance $\sigma^2$. A Gaussian $P_\xi$ gives Gaussian predictive distributions, whereas, for example, a standardized Student-$t$ $P_\xi$ gives heavy-tailed ones.
Starting from $\bm\beta_0$ and a positive-definite matrix $\bm V_0$, the responses are generated sequentially as
\begin{equation}
    Y_1 \mid \bm x_1 \sim P_0(\cdot),
    \qquad\text{and}\qquad
    Y_{n+1} \mid y_{1:n}, \bm x_{1:n+1} \sim P_n(\cdot),
    \label{eqn:seq_constr_rls}
\end{equation}
for $n\geq1$, where $P_n(\cdot)=P(\cdot\mid\bm x_{n+1},\bm\beta_n,\sigma_{n+1})\in\mathcal{P}_{\beta}$ for $n\ge0$, $\bm\beta_n$ denotes an estimator of $\bm\beta$ based on the first $n$ observations $(\bm x_{1:n},y_{1:n})$, and $\sigma_{n+1}$ controls the scale of the next predictive distribution.
Together with $\bm X_{n+1}\sim P_X$ independently of $(\bm X_{1:n},Y_{1:n})$, Equation~\eqref{eqn:seq_constr_rls} defines the joint law $\mathds{P}$ of $(\bm X_n,Y_n)_{n\ge1}$, under which $(\bm X_{1:n},Y_{1:n})$ has distribution $\prod_{i=1}^n P_X(\mathrm d\bm x_i)\,P_{i-1}(\mathrm dy_i)$ for every $n\ge1$.
The key modeling decision is the update of $\bm\beta_n$. We use recursive least squares:
\begin{equation}
    \bm\beta_n = \bm\beta_{n-1} + a_n\bm V_{n-1}\bm x_n(y_n-\bm x_n^\top\bm\beta_{n-1}),
    \label{eqn:rls}
\end{equation}
where $a_n=(1+\bm x_n^\top\bm V_{n-1}\bm x_n)^{-1}$ and $\bm V_{n}$ is a $p\times p$ matrix updated as
\begin{equation}
    \bm V_{n} = \bm V_{n-1} - a_n\bm V_{n-1}\bm x_n\bm x_n^\top\bm V_{n-1}.
    \label{eqn:recursive_p}
\end{equation}
The update is the classical recursive least-squares estimator, written through the Sherman--Morrison formula to avoid repeated matrix inversions \citep{plackett1950some}. Its role here is different from its usual role in online estimation: it is the predictive update that defines the posterior mean and, as the next result shows, induces a martingale posterior.

\begin{assumption}[Random design]
\label{as:beta_sigma}
The covariates satisfy $\bm X_{n+1}\sim P_X$ independently of $(\bm X_{1:n},Y_{1:n})$, for $n\ge0$, so that $(\bm X_n,Y_n)_{n\ge1}$ has probability law $\mathds{P}$.
The sequence $(\bm\beta_n)_{n\ge1}$ is defined by~\eqref{eqn:rls}, $\bm\beta_0$ is fixed, and $\bm V_0$ is fixed and positive-definite.
The predictive scales are $\sigma_{n+1}=s_n(\bm X_{n+1})$, where $s_n:\mathds{R}^p\to(0,\sigma_*]$ is a measurable function of $\bm x$ and $\bm X_{1:n}$, for some $\sigma_*<\infty$, and, $\mathds{P}$-almost surely, $s_n(\bm x)\to\sigma\in(0,\infty)$ for $P_X$-almost every $\bm x$.
\end{assumption}

\begin{theorem}
\label{th:rls}
Under Assumptions~\ref{as:predictive} and~\ref{as:beta_sigma},
\begin{itemize}
\item[$(a)$] the sequence of random vectors $(\bm\beta_n)_{n\ge1}$ converges $\mathds{P}$-almost surely and in $L_2$ to a random limit $\Tilde{\bm\beta}$, and for every $n\ge1$, $\mathds{E}(\Tilde{\bm\beta}\mid Y_{1:n},\bm X_{1:n})=\bm\beta_n$;
\item[$(b)$] the sequence $(\bm X_n,Y_n)_{n\ge1}$ is asymptotically exchangeable with
directing random probability measure
$\tilde P_{\bm X,Y}(\mathrm d\bm x,\mathrm dy)=P_X(\mathrm d\bm x)\,P(\mathrm dy\mid\bm x,\tilde{\bm\beta},\sigma)$;
in particular, the conditional distribution of $Y$ given $\bm X=\bm x$ under $\tilde P_{\bm X,Y}$ is $P(\cdot\mid\bm x,\tilde{\bm\beta},\sigma)$.
\end{itemize}
\end{theorem}

The theorem provides the Bayesian interpretation of the construction. 
The law of $\Tilde{\bm\beta}$ is the implicit prior induced by the predictive update, and the conditional law of $\Tilde{\bm\beta}$ given the observed data is the martingale posterior. Its mean is exactly the \textsc{rls} estimate $\bm\beta_n$. 
Although the law of $(\bm X_n,Y_n)_{n\ge1}$ is in general not exchangeable, but only asymptotically so, the estimator $\bm\beta_n$ does not depend on the ordering of the observations.

\begin{remark}
\label{rem:perm_invariance}
For every $n\ge1$, the recursive least squares estimator $\bm\beta_n$ is invariant to permutations of the observations $(\bm x_1,y_1),\ldots,(\bm x_n,y_n)$. Indeed, \eqref{eqn:rls} and \eqref{eqn:recursive_p} can be rewritten as
\begin{equation}
    \bm\beta_n = \bm V_n \bigg(\bm V_0^{-1}\bm\beta_0 + \sum_{i=1}^n y_i\bm x_i\bigg),
    \qquad\text{and}\qquad
    \bm V_n = \bigg(\bm V_0^{-1} + \sum_{i=1}^n \bm x_i\bm x_i^\top \bigg)^{-1},
\label{eqn:rls_invariance}
\end{equation}
respectively, which are symmetric functions of $(\bm x_1,y_1),\ldots,(\bm x_n,y_n)$.
\end{remark}

Equation~\eqref{eqn:rls_invariance} identifies the finite-sample posterior mean as a generalized ridge estimator. In particular, setting $\bm\beta_0=\bm0$ and $\bm V_0=\kappa\bm I_p$ gives ridge regression, with ordinary least squares recovered as $\kappa\to\infty$ when $\sum_{i=1}^n\bm x_i\bm x_i^\top$ is invertible.
Equation~\eqref{eqn:rls_invariance} also coincides with the posterior mean of a Gaussian linear model with variance $\sigma^2$ and a Gaussian prior on the regression coefficients having mean $\bm\beta_0$ and covariance matrix $\sigma^2\bm V_0$.
The analogy with the posterior mean under a normal linear model is useful for interpreting $\bm\beta_0$ and $\bm V_0$, but the present construction is not that Bayesian model: the predictive distribution may be non-Gaussian, and even a Gaussian plug-in predictive distribution does not in general reproduce the ordinary Gaussian--Gaussian posterior, which is recovered only as a special case (see Section~\ref{sec:asymptotics}).

In many applications, the covariates are not regarded as random, either because they are set by design or simply because the analysis is conditional on them. The whole covariate sequence $\bm x_{1:\infty}$ is then treated as fixed, and Equation~\eqref{eqn:seq_constr_rls} defines a law $\mathds{P}_{\bm x_{1:\infty}}$ of $(Y_n)_{n\ge1}$. The two designs are naturally connected, since $\mathds{P}_{\bm x_{1:\infty}}$ is the conditional law of $(Y_n)_{n\ge1}$ under $\mathds{P}$ given $\bm X_{1:\infty}=\bm x_{1:\infty}$.

\begin{assumption}[Fixed design]
\label{as:fixed_cov}
The covariates $\bm x_{1:\infty}$ are a known and bounded sequence, and $(Y_n)_{n\ge1}$ has probability law $\mathds{P}_{\bm x_{1:\infty}}$.
The sequence $(\bm\beta_n)_{n\ge1}$ is defined by~\eqref{eqn:rls}, $\bm\beta_0$ is fixed, and $\bm V_0$ is fixed and positive-definite.
The predictive scales are $\sigma_n=s_{n-1}(\bm x_n)$, where $s_{n-1}:\mathds{R}^p\to(0,\sigma_*]$ depends only on $\bm x_{1:n-1}$, for some $\sigma_*<\infty$, and $\sigma_n\to\sigma\in(0,\infty)$.
Moreover, we assume that $\bm\Omega:=\lim_{n\to\infty}n^{-1}\sum_{i=1}^n\bm x_i\bm x_i^\top$ exists and is positive definite.
\end{assumption}
\begin{proposition}
\label{prop:rls_fixed}
Under Assumptions~\ref{as:predictive} and~\ref{as:fixed_cov},
\begin{itemize}
\item[$(a)$] the sequence $(\bm\beta_n)_{n\ge1}$ converges $\mathds{P}_{\bm x_{1:\infty}}$-almost surely and in $L_2$ to a random limit $\tilde{\bm\beta}$, and for every $n\ge1$, $\mathds{E}(\tilde{\bm\beta}\mid y_{1:n})=\bm\beta_n$;
\item[$(b)$] $\mathds{P}_{\bm x_{1:\infty}}$-almost surely, the predictive distributions $P_n$ and $P(\cdot\mid\bm x_{n+1},\tilde{\bm\beta},\sigma)$ merge weakly, that is, $\int f\,\mathrm dP_n-\int f\,\mathrm dP(\cdot\mid\bm x_{n+1},\tilde{\bm\beta},\sigma)\to0$ as $n\to\infty$ for every bounded continuous~$f$.
\end{itemize}
\end{proposition}

Part $(b)$ is the fixed-design counterpart of Theorem~\ref{th:rls}$(b)$. Since $\bm x_{n+1}$ changes with~$n$, $P_n$ need not converge, and convergence is replaced by weak merging, a notion from \citet{daristotile1988merging}, weaker than the merging in total variation of \citet{blackwell1962merging}. Informally, for large $n$, $P_n$ is approximately~$P(\cdot\mid\bm x_{n+1},\tilde{\bm\beta},\sigma)$.

Under Assumption~\ref{as:fixed_cov}, the predictive residuals are $Y_i-\bm x_i^\top\bm\beta_{i-1}=\sigma_i\xi_i$, where, by Assumption~\ref{as:predictive}, the $\xi_i$ are iid from $P_\xi$, each independent of~$y_{1:i-1}$. Hence, by~\eqref{eqn:rls} and Proposition~\ref{prop:rls_fixed},
\begin{equation}
    \tilde{\bm\beta}=\bm\beta_n+\sum_{i>n}\bm w_i\,\xi_i,
    \qquad
    \bm w_i=\sigma_ia_i\bm V_{i-1}\bm x_i,
    \label{eqn:representation}
\end{equation}
where the weights $\bm w_i$ are deterministic and the series converges almost surely and in $L_2$.
Since $\xi_{n+1},\xi_{n+2},\ldots$ are independent of $y_{1:n}$, the martingale posterior is a location family centered at $\bm\beta_n$, whose shape does not depend on $y_{1:n}$.

As for the choice of the sequence $(\sigma_n)_{n\ge1}$, one could simply set $\sigma_n=\sigma$ for all $n\ge1$. A better default, however, is $\sigma_n^2=\sigma^2/a_n = \sigma^2(1 + \bm x_n^\top\bm V_{n-1}\bm x_n)$, which is the predictive variance of the conjugate Gaussian linear model with prior covariance matrix $\sigma^2\bm V_0$ and makes the posterior covariance matrix equal to $\sigma^2\bm V_n$ (see Section~\ref{sec:variance}); it satisfies Assumption~\ref{as:beta_sigma} when $P_X$ has bounded support and $\mathds{E}_{P_X}(\bm X\bm X^\top)$ is positive definite, and Assumption~\ref{as:fixed_cov} for any bounded design with $\bm\Omega$ positive definite. The scale $\sigma$ can then be estimated from the data or assigned a prior; see Section~\ref{sec:loo_pll} for further details.
A pipeline for selecting the most appropriate $P_n$ in applications is illustrated in Appendix~C.

\subsection{Posterior variance}
\label{sec:variance}

While Theorem~\ref{th:rls} and Proposition~\ref{prop:rls_fixed} identify the posterior mean for any choice of $P_\xi$, we now characterize the posterior variance under the fixed design of Assumption~\ref{as:fixed_cov}.

\begin{proposition}
\label{prop:variance}
For every $n\ge0$, let $\bm\Sigma_n := \sum_{i>n}\bm w_i\bm w_i^\top = \sum_{i>n} \sigma_i^2 a_i (\bm V_{i-1}-\bm V_i)$, with $\bm w_i=\sigma_ia_i\bm V_{i-1}\bm x_i$.
Under Assumptions~\ref{as:predictive} and~\ref{as:fixed_cov}, for any $n\ge1$,
\begin{equation*}
    \var(\tilde{\bm\beta}\mid y_{1:n}) = \bm\Sigma_n,\qquad
    \mathrm{and}\qquad
    \var(\tilde{\bm\beta}) = \bm\Sigma_0.
\end{equation*}
In particular, if $\sigma_n^2 = \sigma^2/a_n = \sigma^2(1+\bm x_n^\top\bm V_{n-1}\bm x_n)$, then $\bm\Sigma_n = \sigma^2\bm V_n$ and $\bm\Sigma_0 = \sigma^2\bm V_0$.
\end{proposition}

Thus, $\bm\Sigma_n=\sum_{i>n}\sigma_i^2a_i(\bm V_{i-1}-\bm V_i)$ is the exact posterior variance for any choice of $P_\xi$, whereas $\bm\Sigma_0$ is the exact prior variance. Both quantities depend on past and future covariates.
For $\sigma_n^2=\sigma^2/a_n$, the posterior variance of a traditional Bayesian linear model is recovered, and $\bm\Sigma_n=\sigma^2\bm V_n$ depends only on the observed covariates; this is the only choice of $\sigma_n$ for which $\bm\Sigma_n=\sigma^2\bm V_n$ for every $n$.

For a general choice of $\sigma_n$, Proposition~\ref{prop:variance}
may seem to contradict the interpretation of $\sigma^2\bm V_0$ as a prior guess for the covariance of $\tilde{\bm\beta}$, and raises the question of how $\bm V_n$ and $\bm\Sigma_n$
are related.
This is clarified by the following proposition.
\begin{proposition}
\label{prop:convergence_P}
Under Assumptions~\ref{as:predictive} and~\ref{as:fixed_cov}, $n\bm V_n\to\bm\Omega^{-1}$ and $n\bm \Sigma_n\to\sigma^2\bm\Omega^{-1}$.
\end{proposition}
By Proposition~\ref{prop:convergence_P}, both $n\bm\Sigma_n$ and $n\sigma^2\bm V_n$ converge to $\sigma^2\bm\Omega^{-1}$. Hence, $\sigma^2\bm V_n$, which depends only on the observed covariates, is asymptotically equivalent to the posterior covariance for any choice of $\sigma_n$.
The result follows naturally from \eqref{eqn:rls_invariance}, which writes $\bm V_n$ as the inverse of a regularized empirical second-moment matrix.
Lemma~A.3 in the Appendix, on which the proof of Proposition~\ref{prop:convergence_P} is based, also shows that, if $\sigma_n=\sigma$ for all $n\ge1$, then $\bm\Sigma_n\preceq\sigma^2\bm V_n$, where $\bm A\preceq\bm B$ means that $\bm B-\bm A$ is positive semidefinite.
Hence, with a constant predictive variance, the posterior covariance never exceeds $\sigma^2\bm V_n$, obtained with $\sigma_n^2=\sigma^2/a_n$.

\subsection{Asymptotic and exact distributions of $\tilde{\bm\beta}$}
\label{sec:asymptotics}

Building on the exact variance $\bm\Sigma_n$ and its large-sample approximation $\sigma^2\bm V_n$, the next theorem gives an asymptotic approximation to the martingale posterior.

\begin{theorem}
\label{th:asymptotic}
Under Assumptions~\ref{as:predictive} and~\ref{as:fixed_cov},
\begin{equation*}
    \bm \Sigma_n^{-1/2}(\tilde{\bm\beta}-\bm\beta_n)\overset{d}{\longrightarrow}\bm Z,\qquad
    \sigma^{-1}\bm V_n^{-1/2}(\tilde{\bm\beta}-\bm\beta_n)\overset{d}{\longrightarrow}\bm Z,\qquad
    \bm Z\sim\mathcal{N}_p(\bm 0,\bm I_p),
\end{equation*}
conditionally on $y_{1:n}$, as $n\to\infty$, where $\overset{d}{\longrightarrow}$ denotes convergence in distribution.
\end{theorem}

Consequently, for large $n$,
\begin{equation}
    \tilde{\bm\beta}\mid y_{1:n},\bm x_{1:n}\overset{.}{\sim}\mathcal{N}_p(\bm\beta_n, \sigma^2\bm V_n).
    \label{eqn:asy_approx}
\end{equation}
Approximation~\eqref{eqn:asy_approx} always has the exact posterior mean and, when $\sigma_n^2=\sigma^2/a_n$, also the exact posterior variance; for other choices of $\sigma_n$ the variance is approximate.
Notably, it depends only on observed covariates, even though Theorem~\ref{th:asymptotic} assumes that $\bm x_{1:\infty}$ is known.
Theorem~\ref{th:asymptotic} is a predictive, Bayesian result on the martingale posterior, not on the sampling distribution of $\bm\beta_n$, and gives no frequentist guarantees.
When $P_\xi$ is Gaussian, $\tilde{\bm\beta}-\bm\beta_n=\sum_{i>n}\bm w_i\,\xi_i$ is a series of independent Gaussian vectors, and the exact prior and posterior distributions of $\tilde{\bm\beta}$ can be derived.

\begin{theorem}
\label{th:exact}
Under Assumptions~\ref{as:predictive} and~\ref{as:fixed_cov}, assume also that $P_\xi=\mathcal{N}(0,1)$.
Then,
\begin{equation*}
    \Tilde{\bm\beta} \sim \mathcal{N}_p(\bm\beta_0, \bm\Sigma_0),\qquad\text{and}\qquad
    \Tilde{\bm\beta}\mid y_{1:n} \sim \mathcal{N}_p(\bm\beta_n, \bm\Sigma_n),\quad\text{for any }n\ge1.
\end{equation*}
In particular, if $\sigma_n^2 = \sigma^2/a_n = \sigma^2(1+\bm x_n^\top\bm V_{n-1}\bm x_n)$, then
\begin{equation*}
    \Tilde{\bm\beta} \sim \mathcal{N}_p(\bm\beta_0, \sigma^2\bm V_0),\qquad\text{and}\qquad
    \Tilde{\bm\beta}\mid y_{1:n} \sim \mathcal{N}_p(\bm\beta_n, \sigma^2\bm V_n),\quad\text{for any }n\ge1.
\end{equation*}
\end{theorem}

When $\sigma_n^2 = \sigma^2/a_n$ and $P_\xi$ is Gaussian, the ordinary Gaussian--Gaussian prior and posterior are recovered, because the whole construction then coincides with the conjugate model $\bm\beta\sim\mathcal{N}_p(\bm\beta_0,\sigma^2\bm V_0)$, $Y_i\mid\bm x_i,\bm\beta\overset{\mathrm{ind}}{\sim}\mathcal{N}(\bm x_i^\top\bm\beta,\sigma^2)$.
Indeed, the predictive distribution in~\eqref{eqn:seq_constr_rls} is then $\mathcal{N}\{\bm x_{n+1}^\top\bm\beta_n,\sigma^2(1+\bm x_{n+1}^\top\bm V_n\bm x_{n+1})\}$, which is the posterior predictive distribution of the conjugate model, whose posterior mean and covariance are $\bm\beta_n$ and $\sigma^2\bm V_n$.
By the Ionescu-Tulcea theorem, the two models have the same law of $(Y_n)_{n\ge1}$.
The Gaussian case serves mainly as a reference point: it makes the induced prior and posterior fully explicit, whereas non-Gaussian predictive choices retain the same posterior mean but require predictive resampling to characterize the rest of the posterior.

We finally return to the random design of Assumption~\ref{as:beta_sigma}, with $\bm X_n\overset{\mathrm{iid}}{\sim}P_X$ having bounded support and $\mathds E_{P_X}(\bm X\bm X^\top)$ positive definite, and with $\sigma_n\to\sigma$ almost surely, as for $\sigma_n^2=\sigma^2/a_n$.
Then, Assumption~\ref{as:fixed_cov} holds almost surely given $\bm X_{1:\infty}$, and $\mathds E(\tilde{\bm\beta}\mid y_{1:n},\bm X_{1:\infty})=\bm\beta_n$ does not depend on the future covariates. By the law of total variance, $\var(\tilde{\bm\beta}\mid y_{1:n},\bm x_{1:n})=\mathds E(\bm\Sigma_n\mid\bm x_{1:n})$ for every $n\ge0$, where the expectation is over $\bm X_{n+1},\bm X_{n+2},\ldots\overset{\mathrm{iid}}{\sim}P_X$.
With $\sigma_n^2=\sigma^2/a_n$, this equals $\sigma^2\bm V_n$, so neither the posterior mean nor the posterior covariance depends on $P_X$.
This is the weak design invariance property discussed by \citet{sun2026design}: $P_X$ can be replaced in predictive resampling by any distribution with bounded support and $\mathds E_{P_X}(\bm X\bm X^\top)$ positive definite, also when $n<p$.
If, in addition, $P_\xi$ is Gaussian, the equivalence with the conjugate model extends to the joint law of $(\bm X_n,Y_n)_{n\ge1}$, which is therefore exchangeable. Moreover, since $\bm V_n\to\bm 0$ almost surely, the limit $\tilde{\bm\beta}=\lim_{n\to\infty}\bm\beta_n$ is the regression parameter $\bm\beta$ of the conjugate model, and $\tilde{\bm\beta}\mid\bm X_{1:n},Y_{1:n}\sim\mathcal{N}_p(\bm\beta_n,\sigma^2\bm V_n)$ for any $P_X$ with $\mathds E_{P_X}(\bm X\bm X^\top)$ positive definite, even without bounded support. In this case, the whole martingale posterior, and not only its first two moments, does not depend on $P_X$. With a Gaussian $P_\xi$, the update~\eqref{eqn:rls} is also a special case of the score-based recursion of \citet[Appendix~G]{sun2026design}; here, it is kept for any~$P_\xi$.

\subsection{Posterior sampling}
\label{sec:sampling}

Except for the Gaussian case of Theorem~\ref{th:exact}, the martingale posterior distribution for $\tilde{\bm\beta}$ is not available in closed form.
Posterior sampling is carried out by predictive resampling.
Algorithm~\ref{alg:ps_standard} presents the standard predictive resampling when using recursive least squares.
Starting from $\bm\beta_n$ and $\bm V_n$, for $i=n+1,\ldots,N$, the algorithm performs $B$ independent forward samples, each sequentially generating a new pair $(\bm x_i,y_i)$ and updating $\bm\beta_i$ and $\bm V_i$ using \eqref{eqn:rls}--\eqref{eqn:recursive_p}.
The $B$ paths are independent, so they can be run in parallel and the effective sample size of the draws is exactly $B$, whereas \textsc{mcmc} methods may suffer from poor efficiency.
In practice, $P_X$ is usually unknown and must be replaced by an estimator $\hat{P}_X$; the empirical distribution of the observed $\bm x_i$ is typically a good candidate when $n>p$.

Standard predictive resampling can be further accelerated by using the same covariate sequence $\bm X_{n+1:N}$ across all $B$ sample paths, as also done by \citet{sun2026design} and detailed in Algorithm~\ref{alg:ps_fixed}.
This is the fixed-design setting of Section~\ref{sec:model}, with all future covariates drawn before forward sampling begins.
Given the covariates, \eqref{eqn:rls} gives $\bm\beta_N=\bm\beta_n+\sum_{i=n+1}^N\bm w_i\,\xi_i$, the series in~\eqref{eqn:representation} truncated at $N$, with $\xi_{n+1},\ldots,\xi_N$ iid from $P_\xi$.
The weights $\bm w_i=\sigma_ia_i\bm V_{i-1}\bm X_i$ depend only on the covariates, since $\sigma_i=s_{i-1}(\bm X_i)$. They can therefore be computed once, and each draw only requires sampling $\xi_{n+1},\ldots,\xi_N$ and computing a weighted sum.
The effect of stopping at $N$ can also be checked directly: given the covariates, $\var(\bm\beta_N\mid y_{1:n})=\sum_{i=n+1}^N\bm w_i\bm w_i^\top=\bm\Sigma_n-\bm\Sigma_N$. With $\sigma_n^2=\sigma^2/a_n$, this is $\sigma^2(\bm V_n-\bm V_N)$, against the exact $\sigma^2\bm V_n$, so the omitted variance $\sigma^2\bm V_N$ is a by-product of Algorithm~\ref{alg:ps_fixed}.
When $\bm V_0^{-1}$ is negligible, $\bm V_N$ is roughly $n/N$ times $\bm V_n$.
Although sharing the covariates might appear to understate uncertainty in the covariate space, we show in Section~\ref{sec:sim_gaussian} that Algorithms~\ref{alg:ps_standard} and~\ref{alg:ps_fixed} produce approximately the same posterior distribution, with Algorithm~\ref{alg:ps_fixed} being massively faster.

The omitted part is then added back, as in the hybrid algorithm of \citet[Section 3.4]{fong2024asymptotics}. By Theorem~\ref{th:asymptotic} applied at~$N$, $\tilde{\bm\beta}-\bm\beta_N$ is approximately Gaussian with covariance $\bm\Sigma_N$, equal to $\sigma^2\bm V_N$ if $\sigma_n^2=\sigma^2/a_n$, and close to it if $\sigma_n=\sigma$ (Appendix~E), so the last step of Algorithms~\ref{alg:ps_standard} and~\ref{alg:ps_fixed} adds to each draw an independent draw from $\mathcal{N}_p(\bm 0,\sigma^2\bm V_N)$. Thus, with $\sigma_n^2=\sigma^2/a_n$, the draws have mean $\bm\beta_n$ and covariance $\sigma^2\bm V_n$ exactly, for any~$N$. If only these two moments are of interest, they are available in closed form; predictive resampling is needed for the other features of the martingale posterior, such as quantiles, credible intervals and nonlinear functionals. We use $N=10n$ throughout: in all our examples, the resulting credible intervals are close to those obtained with a much larger~$N$ (Appendix~E).

\begin{figure}[t]
\centering
\begin{minipage}[t]{0.48\textwidth}
\algrenewcommand\algorithmicindent{0.6em}% rientro ridotto, solo in questa minipage
\begin{algorithm}[H]
\caption{\small Standard predictive resampling for \textsc{rls}}
\label{alg:ps_standard}
\begin{algorithmic}
\footnotesize
\State Compute $(\bm\beta_n,\bm V_n)$ from the observed data using \eqref{eqn:rls}--\eqref{eqn:recursive_p}
\State Choose $N\gg n$ and the number of draws $B$
\For {$b = 1,\ldots,B$}
\State Set $\bm\beta_n^{(b)}=\bm\beta_n$ and $\bm V_n^{(b)}=\bm V_n$
    \For{$i=n+1,\ldots,N$}
        \State Sample $\bm X_i^{(b)}\sim P_X$
        \State Sample ${Y_i^{(b)}\mid\bm X_i^{(b)}\sim P(\cdot\mid\bm X_i^{(b)},\bm\beta_{i-1}^{(b)},\sigma_i)}$
        \State Update ${a_i^{(b)}=\{1+(\bm X_i^{(b)})^\top\bm V_{i-1}^{(b)}\bm X_i^{(b)}\}^{-1}}$
        \State Update $\bm\beta_i^{(b)}$ and $\bm V_i^{(b)}$ using \eqref{eqn:rls}--\eqref{eqn:recursive_p}
    \EndFor
    \State Sample $\bm Z^{(b)}\sim\mathcal{N}_p(\bm 0,\bm I_p)$ and set
    \State \hspace{\algorithmicindent}$\bm\beta_N^{(b)}\gets\bm\beta_N^{(b)}+\sigma(\bm V_N^{(b)})^{1/2}\bm Z^{(b)}$
\EndFor
\State \Return $\bm\beta_N^{(1)},\ldots,\bm\beta_N^{(B)}$
\end{algorithmic}
\end{algorithm}
\end{minipage}\hfill
\begin{minipage}[t]{0.48\textwidth}
\algrenewcommand\algorithmicindent{0.6em}% rientro ridotto, solo in questa minipage
\begin{algorithm}[H]
\caption{\small Accelerated predictive resampling for \textsc{rls}}
\label{alg:ps_fixed}
\begin{algorithmic}
\footnotesize
\State Compute $(\bm\beta_n,\bm V_n)$ from the observed data using \eqref{eqn:rls}--\eqref{eqn:recursive_p}
\State Choose $N\gg n$ and the number of draws $B$
\For {$i = n+1,\ldots,N$}
    \State Sample $\bm X_i\sim P_X$
    \State Update ${a_i=(1+\bm X_i^\top\bm V_{i-1}\bm X_i)^{-1}}$
    \State Set $\bm w_i=\sigma_ia_i\bm V_{i-1}\bm X_i$
    \State Update $\bm V_i$ using \eqref{eqn:recursive_p}
\EndFor
\For {$b = 1,\ldots,B$}
    \State Sample $\xi_{n+1}^{(b)},\ldots,\xi_N^{(b)}\overset{\mathrm{iid}}{\sim}P_\xi$
    \State Set $\bm\beta_N^{(b)}=\bm\beta_n+\sum_{i=n+1}^N\bm w_i\,\xi_i^{(b)}$
    \State Sample $\bm Z^{(b)}\sim\mathcal{N}_p(\bm 0,\bm I_p)$ and set
    \State \hspace{\algorithmicindent}$\bm\beta_N^{(b)}\gets\bm\beta_N^{(b)}+\sigma\bm V_N^{1/2}\bm Z^{(b)}$
\EndFor
\State \Return $\bm\beta_N^{(1)},\ldots,\bm\beta_N^{(B)}$
\end{algorithmic}
\end{algorithm}
\end{minipage}
\end{figure}

\section{Flexible regression extensions and hyperparameter selection}
\label{sec:extensions}

The predictive framework defined in Section~\ref{sec:rls} is not limited to linear regression. It can be applied whenever the mean of the response is a linear combination of a fixed set of features of the covariates and the coefficients are estimated by penalized least squares.
Each such model is specified by two choices: the features $\bm z_i$, which replace $\bm x_i$ in~\eqref{eqn:rls}--\eqref{eqn:recursive_p}, and the matrix $\bm V_0$, whose inverse acts as a penalty matrix or, equivalently, for which $\sigma^2\bm V_0$ plays the role of a prior covariance matrix, as in the conjugate analogy of Section~\ref{sec:model}.
The rest of the construction of Section~\ref{sec:rls} is unchanged, and the predictive distribution can still be selected separately to reflect the response distribution.
Additive models (Section~\ref{sec:additive_models}) and mixed and Gaussian process models (Section~\ref{sec:mixed_models}) arise in very different applied contexts, but are two instances of this principle. We then discuss a data-driven approach for selecting the hyperparameters (Section~\ref{sec:loo_pll}).

\subsection{Additive models}
\label{sec:additive_models}

In their simplest form, additive models \citep{hastie1986generalized, wood2025generalized}
assume that the mean of the response variable decomposes into a sum of covariate-specific smooth univariate functions, $\mathds{E}(Y\mid\bm x) = \mu + \sum_{j=1}^p f_j(x_j)$, where $x_j$ denotes the $j$th covariate, $f_j(\cdot)$ the corresponding additive component, and $\mu\in\mathds{R}$ the global mean.
Each $f_j$ is commonly estimated through a spline basis expansion, that is, $f_j(x) = \bm h_j(x)^\top\bm\gamma_j$, with $\bm h_j(x)\in\mathds{R}^{r_j}$ containing the evaluations of $r_j$ basis functions at $x$ for component $j$, and $\bm\gamma_j\in\mathds{R}^{r_j}$ the corresponding vector of coefficients.
Since $\mu$ and $f_j$, $j=1,\ldots,p$, are not jointly identifiable, each component is centered so that $\sum_{i=1}^n f_j(x_{i,j})=0$; the resulting basis is kept fixed for future observations.
When penalized splines are employed, the estimation of
$\bm\beta=(\mu, \bm\gamma_1^{\top},\ldots,\bm\gamma_p^{\top})^\top$ from the observations $(y_{1:n},\bm x_{1:n})$ reduces to the minimization of a penalized loss, whose solution is
the penalized least squares estimator
\begin{equation}
    \bm\beta_n = \bm V_n \sum_{i=1}^n y_i\bm z_i,
    \qquad\text{with}\qquad
    \bm V_n^{-1} = \bm \Lambda + \sum_{i=1}^n \bm z_i\bm z_i^\top,
\label{eqn:rls_inv_splines}
\end{equation}
where $\bm z_i=[1, \bm h_1(x_{i,1})^\top, \ldots, \bm h_p(x_{i,p})^\top]^\top\in\mathds{R}^r$,
$\bm x_i=(x_{i,1},\ldots,x_{i,p})$, $r = 1+\sum_{j=1}^p r_j$, and $\bm\Lambda$ is a block-diagonal matrix defined as $\bm\Lambda=\mathrm{diag}(\lambda_\mu,\lambda_1\bm\Lambda_1,\ldots,\lambda_p\bm\Lambda_p)$, where $\lambda_\mu\ge0$ penalizes the global mean ($\lambda_\mu=0$ in our implementation), $\lambda_j>0$ are smoothing parameters and $\bm\Lambda_j\in\mathds{R}^{r_j\times r_j}$ are component-specific penalty matrices, $j=1,\ldots,p$.
The penalty matrix $\bm\Lambda$ is singular in general, so $\bm\Lambda^{-1}$ may not exist and cannot be interpreted as a prior covariance matrix. We therefore set $\bm V_0=(\bm\Lambda+\varepsilon\bm I_r)^{-1}$ with a small $\varepsilon>0$ ($\varepsilon=10^{-10}$ in our implementation), which recovers~\eqref{eqn:rls_inv_splines} as $\varepsilon\to0$.
Thus, the features are the spline evaluations $\bm z_i$ and $\bm V_0$ is a regularized inverse of the penalty matrix: the penalized least squares estimator 
$\bm\beta_n=(\mu_n, \bm\gamma_{n,1}^{\top},\ldots,\bm\gamma_{n,p}^{\top})^\top$
in~\eqref{eqn:rls_inv_splines} is then obtained from~\eqref{eqn:rls}--\eqref{eqn:recursive_p} with $\bm\beta_0=\bm 0$ and $\bm x_i$ replaced by $\bm z_i$, and admits the Bayesian interpretation of Section~\ref{sec:rls}.

\begin{example}[Penalized regression splines]
Low-rank thin plate regression splines \citep{wood2025generalized} are obtained with $\bm h_j$ the thin plate regression spline basis with $r_j\ll n$ functions and $[\bm\Lambda_j]_{\ell,\ell'}=\int h_{j,\ell}''(x)\,h_{j,\ell'}''(x)\,\mathrm{d}x$, $\ell,\ell'=1,\ldots,r_j$.
\end{example}

\begin{example}[P-splines]
P-splines \citep{eilers1996flexible} are obtained with $\bm h_j$ the B-splines of degree $q$ on equally spaced knots and $\bm \Lambda_j=\bm D_{d_j}^\top\bm D_{d_j}$, where $\bm D_{d_j}\in\mathds{R}^{(r_j-d_j)\times r_j}$ is the $d_j$th order difference matrix.
\end{example}

\subsection{Mixed models and Gaussian process regression}
\label{sec:mixed_models}

Suppose that the observations are divided into $q$ groups, and let $g_i\in\{1,\ldots,q\}$ denote the group of observation $i$. A random intercept model assumes
\begin{equation*}
    \mathds{E}(Y_i\mid\bm x_i)=\bm c_i^\top\bm\eta+\alpha_{g_i},
\end{equation*}
where $\bm c_i\in\mathds{R}^d$ are individual-specific covariates with effects $\bm\eta$ and $\alpha_1,\ldots,\alpha_q$ are group-specific intercepts; a global intercept, or a constant in $\bm c_i$, would be collinear with the group indicators.
This mean is linear in $\bm\beta=(\bm\eta^\top,\alpha_1,\ldots,\alpha_q)^\top$, with features $\bm z_i=(\bm c_i^\top,\bm m_i^\top)^\top$, where $\bm m_i\in\{0,1\}^q$ has $j$th entry equal to $1$ if $g_i=j$ and $0$ otherwise.
Hence, the \textsc{rls} estimator $\bm\beta_n$ contains the estimates of the covariate effects and of the group-specific intercepts, and the type of model is determined by the choice of $\bm V_0=\mathrm{blockdiag}(\bm D,\bm K)$, where $\bm D\in\mathds{R}^{d\times d}$ and $\bm K\in\mathds{R}^{q\times q}$ are positive definite and play the role of prior covariance matrices, up to $\sigma^2$, of $\bm\eta$ and of the group-specific intercepts, respectively. A common choice is $\bm D=\mathrm{diag}(v_1,\ldots,v_d)$, but any positive definite $\bm D$ can be used.

Setting $\bm K=\mathrm{diag}(v_{d+1},\ldots,v_{d+q})$ yields a model with group-specific fixed intercepts, whereas $\bm K=\tau\bm I_q+v_\mu\bm 1_q\bm 1_q^\top$ gives the corresponding random intercept model, with $\alpha_j=\mu+u_j$ for a global intercept~$\mu$ and independent group effects~$u_j$, where $\bm 1_q$ is the vector of ones and $\tau$ is either selected by maximizing the \textsc{loo-pll} or assigned a prior, with the prequential log-likelihood used as a likelihood; see Section~\ref{sec:loo_pll}.
A non-diagonal $\bm K$ allows the group-specific intercepts to be correlated. In particular, when the groups are $q$ fixed locations with coordinates $\bm s_1,\ldots,\bm s_q$ and $\bm K$ is a kernel covariance matrix, with entries $k(\bm s_j,\bm s_\ell)$ for a kernel function $k$, we obtain Gaussian process regression at fixed locations, and $\alpha_1,\ldots,\alpha_q$ are the spatial effects.

Under a Gaussian likelihood, all these models are easily estimated within the traditional Bayesian framework. Under a non-Gaussian one, estimation can become much more expensive, whereas here the posterior mean $\bm\beta_n$ is unchanged and only the predictive distribution $P_n$ needs to be adapted.

\subsection{Hyperparameter selection}
\label{sec:loo_pll}

As discussed in the previous sections, the \textsc{rls} estimator may depend on hyperparameters contained in $\bm V_0$, while predictive resampling depends on the scale and shape parameters of $P_n$.
It is therefore desirable to select these hyperparameters in a data-driven way.
A common approach in predictive Bayes methods \citep{fong_holmes_2023, fong2024asymptotics, agnoletto2025nonparametric} is to rely on the so-called \textit{prequential log-likelihood} \citep{dawid1984present}, $\sum_{i=1}^n\log p_{i-1}(y_i\mid \bm x_i)$, which can be interpreted as a log-likelihood arising from a sequential construction.
The prequential log-likelihood can then either be maximized to select the hyperparameters or, after placing a prior on them, be used as a likelihood in a fully Bayesian setting.

However, this quantity has the important drawback of depending on the ordering of the observations, so it should be averaged over several permutations to heuristically ensure invariance to the ordering.
Moreover, we have observed in practice that it can be numerically unstable when maximized.
Therefore, rather than averaging over permutations, which increases the computational burden, we propose using the leave-one-out prequential log-likelihood (\textsc{loo-pll}), defined as
\begin{equation*}
    \sum_{i=1}^n \log p(y_i\mid\bm x_i, \bm\beta_{-i}, \sigma_{(i)}),
\end{equation*}
where $p(\cdot)$ denotes the density function of the distribution in $\mathcal{P}_\beta$, $\bm\beta_{-i}$ is the estimate of the regression coefficients based on all observations except the $i$th, and $\sigma_{(i)}$ is the predictive scale of the $i$th observation when it comes last, e.g.\ $\sigma_{(i)}^2=\sigma^2/(1-\bm x_i^\top\bm V_n\bm x_i)$ when $\sigma_n^2=\sigma^2/a_n$.
Up to a factor $n$, this is equivalent to evaluating the prequential log-likelihood contribution of the last observation based on the coefficients estimated from the other $n-1$ observations and averaging it over all possible permutations of the data, since each observation comes last in $(n-1)!$ of the $n!$ permutations and, by Remark~\ref{rem:perm_invariance}, $\bm\beta_{-i}$ does not depend on the ordering of the other observations. An analogous identity relates the marginal likelihood to exhaustive cross-validation \citep{fong2020marginal}.
The \textsc{loo-pll} is therefore invariant to the ordering of the observations.
It can also be computed efficiently, since the leave-one-out predictions satisfy $\bm x_i^\top\bm\beta_{-i}= y_i - (y_i-\bm x_i^\top\bm\beta_n)/(1-\bm x_i^\top\bm V_n\bm x_i)$ for $i=1,\ldots,n$, so $\bm\beta_n$ and $\bm V_n$ need to be computed only once.
The \textsc{loo-pll} can thus replace the standard prequential log-likelihood for hyperparameter selection, and in our experience it performs better in practice.

\section{Simulation study}
\label{sec:sim}

\subsection{Comparison between uncertainty quantification methods}
\label{sec:sim_gaussian}

We first check predictive resampling in a linear regression setting in which the exact posterior is available.
The response is Gaussian with known standard deviation $\sigma=2$, the $p=50$ covariates are standard multivariate normal, and we consider two scenarios, $n=3p$ and $n=10p$, each with $30$ simulated datasets; the true coefficients are drawn once from $\mathcal{N}_p(\bm 0,4\bm I_p)$.
We set $\bm\beta_0=\bm 0$, $\bm V_0=1000\bm I_p$, $\hat P_X$ equal to the empirical distribution of the observed covariates, $B=10{,}000$ and $N=10n$, and consider both $\sigma_n^2=\sigma^2/a_n$ and $\sigma_n^2=\sigma^2$.
For each choice, we compare Algorithm~\ref{alg:ps_standard} (\textsc{ps-1}) and Algorithm~\ref{alg:ps_fixed} (\textsc{ps-2}) with $B$ independent draws from the exact posterior $\mathcal{N}_p(\bm\beta_n,\bm\Sigma_n)$ of Theorem~\ref{th:exact} (\textsc{exact}), where, if $\sigma_n^2=\sigma^2$, $\bm\Sigma_n$ is computed given a sequence of future covariates drawn from $\hat P_X$, whose first $N-n$ elements are those of \textsc{ps-2}.
We report the errors of the sample mean $\hat{\bm\beta}$ and covariance $\hat{\bm\Sigma}$ of the draws, the average width and the frequentist coverage of the 95\% credible intervals for the coefficients, and the computing time on an Apple \textsc{m2} chip with 8 cores, with an \textsc{r/c++} implementation.
An analogous study for heavy-tailed data is reported in Appendix~B.1.

\begin{table}[t]
\centering
\caption{\small Comparison of uncertainty quantification from predictive resampling, with Algorithms~\ref{alg:ps_standard} (\textsc{ps-1}) and~\ref{alg:ps_fixed} (\textsc{ps-2}), with $B$ draws from the exact posterior $\mathcal{N}_p(\bm\beta_n,\bm\Sigma_n)$ (\textsc{exact}): logarithm of the mean squared error of the posterior mean, $\log(\|\hat{\bm\beta}-\bm\beta_n\|^2/p)$, logarithm of the Frobenius norm of the lower-triangular part of $\sigma^{-2}(\hat{\bm\Sigma}-\bm\Sigma_n)$, average width and coverage of the 95\% intervals for the $p$ regression coefficients, and computation time in seconds. For \textsc{exact}, the errors are Monte Carlo errors and the intervals are the exact Gaussian ones. Values are averages over the simulated datasets, with standard deviations in parentheses.}
\vspace{0.2cm}
\label{tab:sim_gaussian}
\footnotesize
\begin{tabular}{l l ccccc}
\hline
& & $\log(\|\hat{\bm\beta}-\bm\beta_n\|^2/p)$ & $\log\|\hat{\bm\Sigma}-\bm\Sigma_n\|_F$ & \textsc{width} & \textsc{cover} & \textsc{time}\\
\hline
\multicolumn{7}{l}{$n=3p$}\\
$\sigma_n^2=\sigma^2a_n^{-1}$ & \textsc{ps-1} & $-12.42$ & $-5.61$ & $0.780$ & $0.955$ & $6.18$ \\
& & $^{(0.19)}$ & $^{(0.04)}$ & $^{(0.009)}$ & $^{(0.029)}$ & $^{(0.28)}$ \\
 & \textsc{ps-2} & $-12.41$ & $-5.61$ & $0.779$ & $0.949$ & $0.08$ \\
& & $^{(0.27)}$ & $^{(0.04)}$ & $^{(0.009)}$ & $^{(0.034)}$ & $^{(0.00)}$ \\
 & \textsc{exact} & $-12.42$ & $-5.61$ & $0.782$ & $0.956$ & $-$ \\
& & $^{(0.26)}$ & $^{(0.03)}$ & $^{(0.009)}$ & $^{(0.027)}$ &  \\
\cmidrule{2-7}
$\sigma_n^2=\sigma^2$ & \textsc{ps-1} & $-12.57$ & $-5.73$ & $0.723$ & $0.936$ & $6.33$ \\
& & $^{(0.19)}$ & $^{(0.04)}$ & $^{(0.009)}$ & $^{(0.039)}$ & $^{(0.07)}$ \\
 & \textsc{ps-2} & $-12.57$ & $-5.76$ & $0.722$ & $0.935$ & $0.08$ \\
& & $^{(0.27)}$ & $^{(0.04)}$ & $^{(0.009)}$ & $^{(0.039)}$ & $^{(0.00)}$ \\
 & \textsc{exact} & $-12.57$ & $-5.76$ & $0.724$ & $0.935$ & $-$ \\
& & $^{(0.26)}$ & $^{(0.03)}$ & $^{(0.008)}$ & $^{(0.037)}$ &  \\
\hline
\multicolumn{7}{l}{$n=10p$}\\
$\sigma_n^2=\sigma^2a_n^{-1}$ & \textsc{ps-1} & $-13.93$ & $-7.11$ & $0.369$ & $0.946$ & $21.57$ \\
& & $^{(0.22)}$ & $^{(0.03)}$ & $^{(0.002)}$ & $^{(0.034)}$ & $^{(0.87)}$ \\
 & \textsc{ps-2} & $-13.92$ & $-7.12$ & $0.368$ & $0.950$ & $0.24$ \\
& & $^{(0.22)}$ & $^{(0.03)}$ & $^{(0.001)}$ & $^{(0.029)}$ & $^{(0.01)}$ \\
 & \textsc{exact} & $-13.94$ & $-7.11$ & $0.370$ & $0.947$ & $-$ \\
& & $^{(0.21)}$ & $^{(0.02)}$ & $^{(0.001)}$ & $^{(0.033)}$ &  \\
\cmidrule{2-7}
$\sigma_n^2=\sigma^2$ & \textsc{ps-1} & $-13.97$ & $-7.16$ & $0.360$ & $0.941$ & $21.45$ \\
& & $^{(0.21)}$ & $^{(0.03)}$ & $^{(0.002)}$ & $^{(0.034)}$ & $^{(0.68)}$ \\
 & \textsc{ps-2} & $-13.97$ & $-7.17$ & $0.360$ & $0.941$ & $0.26$ \\
& & $^{(0.22)}$ & $^{(0.03)}$ & $^{(0.001)}$ & $^{(0.031)}$ & $^{(0.01)}$ \\
 & \textsc{exact} & $-13.99$ & $-7.16$ & $0.361$ & $0.941$ & $-$ \\
& & $^{(0.21)}$ & $^{(0.02)}$ & $^{(0.001)}$ & $^{(0.034)}$ &  \\
\hline
\end{tabular}
\end{table}

The results are reported in Table~\ref{tab:sim_gaussian}.
With the correction, the draws of both algorithms have exact mean and, if $\sigma_n^2=\sigma^2/a_n$, exact covariance, so this experiment mainly checks the implementation: as expected, the errors of \textsc{ps-1} and \textsc{ps-2} are comparable with the Monte Carlo errors of \textsc{exact}, and the widths and coverages essentially coincide with the exact ones.
Less obviously, \textsc{ps-2} performs as \textsc{ps-1} also when $\sigma_n^2=\sigma^2$, where $\bm\Sigma_n$ depends on the future covariates: sharing them across paths has a negligible effect.
With $\sigma_n^2=\sigma^2/a_n$, the coverage is close to the nominal level, as expected since the martingale posterior then coincides with the posterior of the conjugate model; with $\sigma_n^2=\sigma^2$, the intervals are narrower, in agreement with the bound $\bm\Sigma_n\preceq\sigma^2\bm V_n$ of Section~\ref{sec:variance}, and the coverage is slightly below the nominal level.
The correction is not negligible: the omitted part of the posterior variance is about $n/N=10\%$, and without it the intervals would be about $5\%$ narrower.
Finally, \textsc{ps-2} is about 80 times faster than \textsc{ps-1}, and is therefore preferable in practice.

\subsection{Misspecification of the predictive distribution}
\label{sec:ill}

We now illustrate how the choice of $P_n$ affects the martingale posterior of the regression coefficients, through two simulated examples with $n=15$ observations and $p=5$ covariates.
We generate $y_i=\bm x_i^\top\bm\beta_0^*+\phi_0\xi_i$, $i=1,\ldots,n$, with $\bm x_i$ iid from $\mathcal{N}_p(\bm 0,\bm I_p)$, $\bm\beta_0^*=(-1.54,1.81,-2.21,-0.22,-3.48)^\top$ and $\phi_0=0.2$.
In the first example, the $\xi_i$ are iid from a Student-$t$ distribution with $\nu_0=3$ degrees of freedom; in the second, they are iid from a skew Student-$t$ distribution \citep{azzalini2003distributions} with $\nu_0=3$ and skewness parameter $\alpha_0=3$, centered to have zero mean.
In each example, we compare a $P_n$ that matches the error distribution, namely a location-scale Student-$t$ and a skew Student-$t$ centered to have mean $\bm x^\top\bm\beta$, respectively, with a Gaussian one, which, with $\sigma_n^2 = \sigma^2 / a_n$, coincides with the conjugate Bayesian linear model.
We set $\bm\beta_0=\bm 0$, $\bm V_0=1000\bm I_p$ and $\sigma_n^2=\sigma^2/a_n$, select the hyperparameters by maximizing the \textsc{loo-pll}, and run Algorithm~\ref{alg:ps_fixed} with $N=10n$, $\hat P_X$ equal to the empirical distribution of the observed covariates, and $B=10{,}000$.
The \textsc{loo-pll} estimates are $\hat\phi=0.21$ and $\hat\nu=2.60$ for the Student-$t$ $P_n$ in the first example, and $\hat\phi=0.26$, $\hat\nu=4.07$ and $\hat\alpha=3.80$ for the skew Student-$t$ $P_n$ in the second; in the first example, the Gaussian $P_n$ inflates the scale, $\hat\phi=0.33$, to match the error variance.

\begin{figure}[t]
\centering
\includegraphics[width=\textwidth]{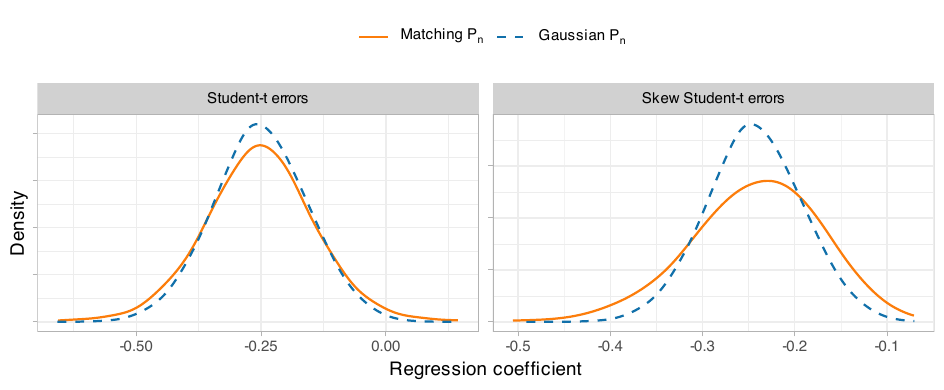}
\caption{\small Posterior distribution of a regression coefficient obtained with a $P_n$ matching the error distribution and with a Gaussian $P_n$, for Student-$t$ errors (left) and skew Student-$t$ errors (right).}
\label{fig:ill_post}
\end{figure}

Figure~\ref{fig:ill_post} shows the posterior distribution of the fourth regression coefficient.
With a matching $P_n$, the heavy tails of the errors, and in the second example also their skewness, carry over to the martingale posterior, which is more dispersed than, and in the second example skewed relative to, the posterior of the Gaussian linear model.
The choice of $P_n$ thus affects not only predictive quantities, such as the quantiles of $Y\mid\bm x$, but also inference on the regression coefficients.

\subsection{Additive models with Student-$t$ predictive distribution}
\label{sec:sim_am}
We now compare our approach with well-established Bayesian methods in a nonlinear regression setting with heavy-tailed errors.
We generate $y_i = \mu+f(\bm x_i) + \phi_0\xi_i$, $i=1,\ldots,n$, with covariates iid uniform on $(-2,2)$, $\xi_i$ iid from a Student-$t$ distribution with $\nu_0=5$ degrees of freedom, $\mu=1$, $\phi_0=1$ and $f(\bm x_i) = \sum_{j=1}^p f_j(x_{i,j})$, where $f_j=f^*_{k}$, with $k=1+\{(j-1)\bmod 4\}$, is centered over the sample and
\begin{equation*}
    f^*_1(x) = \frac{1}{2}x^3 - x,\quad
    f^*_2(x) = 3x\exp\big\{-x^2\big\},\quad
    f^*_3(x) = x^2 - 1,\quad
    f^*_4(x) = 2\sin(\pi x).
\end{equation*}
We set $p=15$ and consider two scenarios, $n=30p$ and $n=100p$, each with $30$ simulated datasets.
On a test set of size $10{,}000$, we compute the mean squared error (\textsc{mse}) of the posterior mean of $f(\bm x)$ and the average coverage and width of the 95\% credible intervals for $f(\bm x)$, where the coverage at a test point is the proportion of datasets whose interval contains the true value; we do the same for $Q_{0.90}(\bm{x})$, the $0.90$ quantile of the response at~$\bm{x}$.
We also report the computing time on an Apple \textsc{m2} chip with 8 cores, for the posterior mean alone, including the selection of the hyperparameters, and in total.

Our method (\textsc{rls-$t$}) uses the penalized regression splines of Section~\ref{sec:additive_models} with a location-scale Student-$t$ $P_n$ and $\sigma_n^2=\sigma^2/a_n$, where $\nu>2$ and $\sigma^2=\phi^2\nu(\nu-2)^{-1}$; $\phi$, $\nu$ and the smoothing parameters are selected by maximizing the \textsc{loo-pll}, and Algorithm~\ref{alg:ps_fixed} is run with $N=10n$ and~$B=5{,}000$. We also include its Gaussian version (\textsc{rls-$n$}).
The first competitor is the traditional Bayesian counterpart of our model, with a Student-$t$ likelihood (\textsc{blm}). The coefficients have a Gaussian prior with block-diagonal precision matrix, whose blocks are the spline penalty matrices multiplied by a smoothing parameter and divided by $\phi^2$, improper along the null space of each penalty, and each smoothing parameter has a $\mathrm{Gamma}(2,0.5)$ prior, with shape $2$ and rate $0.5$.
We sample from its posterior with the \textsc{r} package \texttt{brms}, based on \textsc{stan}, running five chains of $2{,}000$ draws and discarding the first $1{,}000$ as burn-in; the sampler reports no divergent transitions.
The second competitor is Bayesian additive regression trees (\textsc{bart}; \citealp{chipman2010bart}), with $55{,}000$ draws, of which the first $50{,}000$ are discarded as burn-in.
The \textsc{mcmc} chain lengths are chosen, when possible, to give an effective sample size (\textsc{ess}) close to~$B$ (Appendix~B.3), and the times exclude the compilation of the \textsc{stan} model.

\begin{table}[t]
\centering
\caption{\small Comparison between our proposed methods, with Student-$t$ (\textsc{rls-$t$}) and Gaussian (\textsc{rls-$n$}) predictive distributions, the traditional Bayesian linear model with Student-$t$ likelihood (\textsc{blm}) and \textsc{bart} in terms of out-of-sample \textsc{mse}, average coverage and width of credible intervals for $f(\bm x)$ and $Q_{0.90}(\bm x)$, and time for obtaining the posterior mean (\textsc{pm}) and total, respectively. Values are averages over the simulated datasets, with standard deviations in parentheses.}
\vspace{0.2cm}
\label{tab:sim}
\footnotesize
\begin{tabular}{ll cccc cc cccc}
\hline
& & \multicolumn{4}{c}{$n=30p$} & & & \multicolumn{4}{c}{$n=100p$}\\
& & {\textsc{rls-$t$}} & {\textsc{rls-$n$}} & {\textsc{blm}} & {\textsc{bart}} & & & {\textsc{rls-$t$}} & {\textsc{rls-$n$}} & {\textsc{blm}} & {\textsc{bart}}\\
\hline
$f(\bm x)$ & \textsc{mse} & $0.438$ & $0.452$ & $\bm{0.420}$ & $1.291$ & & &
$0.119$ & $0.122$ & $\bm{0.100}$ & $0.679$ \\
& & $^{(0.084)}$ & $^{(0.095)}$ & $^{(0.077)}$ & $^{(0.168)}$ & & &
$^{(0.014)}$ & $^{(0.014)}$ & $^{(0.012)}$ & $^{(0.056)}$   \\
& \textsc{cover} & $0.935$ & $0.931$ & $\bm{0.956}$ & $0.975$ & & &
$\bm{0.948}$ & $0.944$ & $0.961$ & $\bm{0.952}$ \\
& & $^{(0.022)}$ & $^{(0.023)}$ & $^{(0.017)}$ & $^{(0.011)}$ & & &
$^{(0.014)}$ & $^{(0.014)}$ & $^{(0.013)}$ & $^{(0.013)}$   \\
& \textsc{width ci} & $2.437$ & $2.432$ & $2.598$ & $5.229$ & & &
$1.331$ & $1.330$ & $1.302$ & $3.336$ \\
& & $^{(0.128)}$ & $^{(0.129)}$ & $^{(0.113)}$ & $^{(0.251)}$ & & &
$^{(0.024)}$ & $^{(0.023)}$ & $^{(0.028)}$ & $^{(0.087)}$   \\
\hline
$Q_{0.90}(\bm x)$ & \textsc{mse} & $\bm{0.447}$ & $0.483$ & $0.507$ & $1.397$ & & &
$0.120$ & $0.147$ & $\bm{0.104}$ & $0.680$ \\
& & $^{(0.083)}$ & $^{(0.110)}$ & $^{(0.111)}$ & $^{(0.183)}$ & & &
$^{(0.014)}$ & $^{(0.019)}$ & $^{(0.012)}$ & $^{(0.056)}$   \\
& \textsc{cover} & $0.933$ & $0.921$ & $\bm{0.935}$ & $0.970$ & & &
$0.946$ & $0.917$ & $0.959$ & $\bm{0.952}$ \\
& & $^{(0.021)}$ & $^{(0.024)}$ & $^{(0.022)}$ & $^{(0.014)}$ & & &
$^{(0.014)}$ & $^{(0.017)}$ & $^{(0.013)}$ & $^{(0.013)}$   \\
& \textsc{width ci} & $2.437$ & $2.432$ & $2.617$ & $5.245$ & & &
$1.331$ & $1.330$ & $1.310$ & $3.339$ \\
& & $^{(0.128)}$ & $^{(0.129)}$ & $^{(0.114)}$ & $^{(0.252)}$ & & &
$^{(0.024)}$ & $^{(0.023)}$ & $^{(0.028)}$ & $^{(0.087)}$   \\
\hline
Time & \textsc{pm} & $0.32$ & $\bm{0.17}$ & $27.50$ & $177.48$ & & &
$0.77$ & $\bm{0.60}$ & $76.70$ & $426.04$ \\
& & $^{(0.09)}$ & $^{(0.04)}$ & $^{(2.38)}$ & $^{(2.65)}$ & & &
$^{(0.19)}$ & $^{(0.09)}$ & $^{(4.64)}$ & $^{(3.75)}$   \\
& \textsc{total} & $0.72$ & $\bm{0.55}$ & $27.50$ & $177.48$ & & &
$2.00$ & $\bm{1.74}$ & $76.70$ & $426.04$ \\
& & $^{(0.10)}$ & $^{(0.05)}$ & $^{(2.38)}$ & $^{(2.65)}$ & & &
$^{(0.51)}$ & $^{(0.09)}$ & $^{(4.64)}$ & $^{(3.75)}$   \\
\hline
\end{tabular}
\end{table}

The results are reported in Table~\ref{tab:sim}.
\textsc{blm} yields the most accurate point estimates, except for $Q_{0.90}(\bm x)$ when $n=30p$, where \textsc{rls-$t$} is the most accurate; overall, \textsc{rls-$t$} is close to \textsc{blm} and much more accurate than \textsc{bart}. This is expected, since the true $f$ is additive, whereas \textsc{bart} also accommodates interactions among the covariates.
The point estimates of \textsc{rls-$t$} and \textsc{rls-$n$} differ only through their estimated smoothing parameters.
The coverage of \textsc{rls-$t$} is slightly below the nominal level, and that of \textsc{blm} slightly above it, except for $Q_{0.90}(\bm x)$ when $n=30p$; \textsc{rls-$n$} has lower coverage for $Q_{0.90}(\bm{x})$, which depends more heavily on the tails of $P_n$, and \textsc{bart} attains coverage at or above the nominal level with intervals more than twice as wide.
Finally, our posterior mean requires no posterior sampling, whereas the competitors must run the full \textsc{mcmc} to obtain it, so for them the two times in Table~\ref{tab:sim} coincide. In both scenarios, \textsc{rls-$t$} and \textsc{rls-$n$} are more than an order of magnitude faster than the competitors, both for the posterior mean and in total time. Appendix~B.2 compares our method also with frequentist additive models fitted with the \textsc{r} package \texttt{mgcv} \citep{wood2011fast}: the Student-$t$ additive model gives more accurate point estimates, while our method is faster, even including predictive resampling.

\section{Analysis of NHANES serum triglycerides dataset}
\label{sec:application}

We analyze fasting serum triglyceride concentrations (mg/dL) from the 2021--2023 cycle of the National Health and Nutrition Examination Survey (\textsc{nhanes}), focusing on US-born males aged 12 to 40 years tested between May and October.
\textsc{nhanes} is a stratified, multistage probability sample of the civilian population of the United States.
Triglycerides are the circulating storage form of dietary and hepatically synthesized fat, transported in chylomicrons and very-low-density lipoproteins (\textsc{vldl}).
Monitoring the level of plasma triglycerides is particularly important because it is a risk-enhancing factor for cardiovascular diseases \citep{hokanson1996plasma, sarwar2007triglycerides}.
Since both triglycerides and glucose respond to recent food intake, the analysis is restricted to the 119 participants who fasted between 12 and 24 hours before the examination.
For each participant, the dataset records eight characteristics: age (years), body mass index (\textsc{bmi}), waist circumference, high-density lipoprotein (\textsc{hdl}) cholesterol, fasting plasma glucose, average systolic blood pressure (\textsc{sbp}),
the ratio of family income to the federal poverty threshold, and ethnicity (Mexican American, other Hispanic, non-Hispanic white, non-Hispanic black, non-Hispanic Asian, and other or multiracial).
Modeling the relation between these covariates and triglyceride concentration is of practical interest: \textsc{bmi}, \textsc{hdl} cholesterol, and fasting glucose are all proxies for insulin resistance, the principal driver of hypertriglyceridemia \citep{sparks2012selective}.

Several relations between the covariates and the logarithm of triglyceride concentration are expected to be nonlinear, thereby motivating modeling the mean through additive smooth components.
Specifically, we include the covariates age, \textsc{bmi}, \textsc{hdl}, plasma glucose, average \textsc{sbp}, and income level via penalized regression splines, while ethnicity is encoded with five dummy indicators, with non-Hispanic white as the baseline.
We exclude waist circumference from the analysis because it is highly correlated with \textsc{bmi} (correlation 0.96 in the training set).
A location-scale Student-$t$ predictive distribution is a natural choice for modeling log-triglyceride concentration, since the distribution is typically heavy-tailed \citep{hegele2014polygenic}.
Additional details on the choice of $P_n$ are contained in Appendixn~C.

We compare the results obtained with our approach against the same model fitted under the traditional Bayesian approach with the \textsc{r} package \texttt{brms}.
We randomly split the dataset into a training set of 83 observations and a test set of 36 observations, used to estimate the models and to evaluate out-of-sample predictive accuracy, respectively.
Out-of-sample accuracy is evaluated through the \textsc{mse} between the posterior mean of $\mathds{E}(Y\mid\bm x)$ and the corresponding observed responses on the test set.
Importantly, we also assess 
the computational efficiency of the posterior sampling algorithm.
For the proposed method, we set $\sigma_n^2=\sigma^2/a_n$, as in Section~\ref{sec:sim_am}, leave the coefficients of the ethnicity indicators unpenalized, as the global mean, and select the degrees of freedom $\nu$, the scale parameter $\phi$, and the smoothing parameters by maximizing the \textsc{loo-pll}, obtaining $\nu = 13.85$ and $\phi = 0.36$.
Uncertainty quantification is carried out via Algorithm~\ref{alg:ps_fixed} with $B=10{,}000$ and $N=10n$. For the traditional Bayesian model, we use the prior of Section~\ref{sec:sim_am}, with a $\mathrm{Gamma}(2,0.5)$ prior on each smoothing parameter, and we run five \textsc{mcmc} chains of $3{,}000$ iterations each and discard the first $1{,}000$ as burn-in; the target acceptance rate of \textsc{stan} is raised from $0.8$ to $0.95$, since the default yields divergent transitions, also with a longer burn-in.

\begin{figure}[t]
\centering
\includegraphics[width=\textwidth]{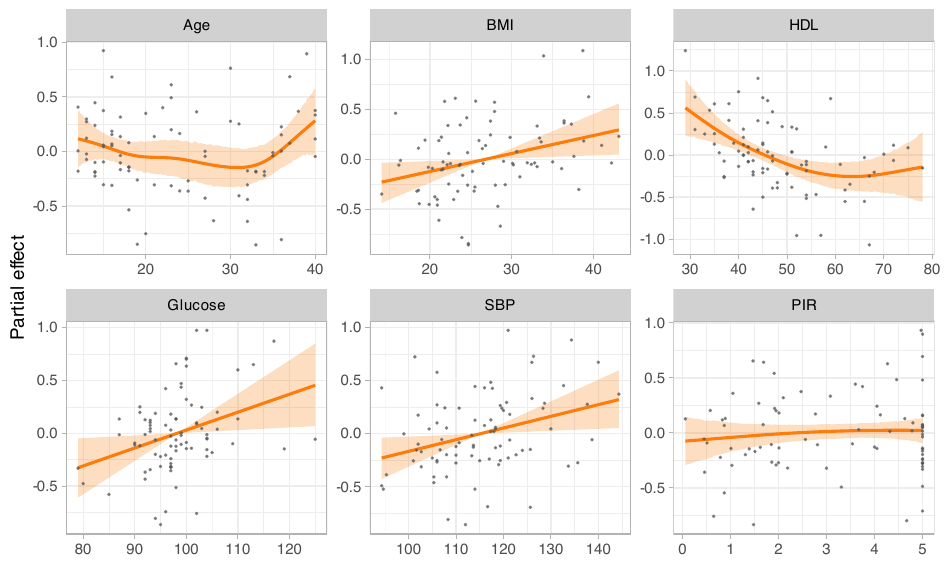}
\caption{\small Estimated covariate-specific effects with pointwise 95\% highest posterior density bands, obtained using our predictive additive model, and partial residuals. The bands are truncated to the range of the partial residuals.}
\label{fig:app_1}
\end{figure}

Our approach obtains an accuracy similar to that of the traditional Bayesian additive model with Student-$t$ likelihood (\textsc{mse} of $0.263$ versus $0.249$). Selecting the hyperparameters and computing the posterior mean take 0.02 seconds and predictive resampling takes 0.09 seconds, whereas \textsc{mcmc} requires 6.68 seconds. In addition, predictive resampling yields posterior draws that are independent by construction, given the shared future covariates, whereas the average \textsc{ess} of the \textsc{mcmc} draws across regression parameters is $5{,}424$ out of $10{,}000$.
The traditional Bayesian model is, however, sensitive to the prior on the smoothing parameters: with a vague $\mathrm{Gamma}(0.001,0.001)$ prior, the sampler reports $235$ divergent transitions out of $10{,}000$ draws, the average \textsc{ess} drops to $1{,}266$, and \textsc{mcmc} requires 39.21 seconds. Our approach requires no such prior, since its hyperparameters are selected by maximizing the \textsc{loo-pll}.

The results obtained with our predictive additive model are shown in Figure~\ref{fig:app_1}.
The relation between \textsc{hdl} and log-triglycerides is clearly nonlinear.
The inferred shape is consistent with a well-documented saturating exchange process 
\citep{patsch1984postprandial, barter2003cholesteryl}.
The effect of age on log-triglycerides is also nonlinear: it decreases up to about age 30 and increases afterwards.
For fasting glucose, \textsc{bmi} and \textsc{sbp}, the estimated effect is linear and positive, which is consistent with the literature
\citep{reaven1988role, adiels2005overproduction}.
The remaining covariates, including ethnicity, show no clear association with log-triglyceride concentration, as their 95\% credible bands and intervals include zero.
In the analyzed sample, age, \textsc{bmi}, \textsc{hdl} cholesterol, fasting glucose and \textsc{sbp} are thus associated with log-triglycerides; the analysis does not use the \textsc{nhanes} survey weights.

\section{Discussion}

Recursive least squares is a popular estimator in the frequentist literature for streaming data. We have proposed a novel interpretation within a predictive Bayes framework, expressing it as a predictive rule that characterizes an implicit prior on the regression coefficients of a linear model. 
The proposed method addresses some limitations of existing predictive Bayes approaches: it provides a new method for regression and, despite assuming that observations are generated sequentially, it is invariant to their ordering.
While in traditional Bayesian models the choice of likelihood and prior is often restricted to computationally convenient combinations, our approach offers greater flexibility, since the posterior mean $\bm\beta_n$ always coincides with the one obtained under a Gaussian--Gaussian model whereas the plug-in predictive distribution $P_n$ may be chosen separately. 
The approach extends to nonlinear additive models, mixed models, and Gaussian process regression through an appropriate choice of the initial $\bm V_0$.

Some limitations remain. First, the approach does not generalize to the full class of generalized linear models, as no closed-form least squares estimator is available in that setting; for logistic regression, one possibility could be to adopt the construction of \citet[Section~5.2]{fortini2024exchangeability} and study how extensions beyond the simple regression case can be accommodated.
Second, $\bm\beta_n$ is the generalized ridge estimator for every choice of $P_\xi$: a heavy-tailed $P_n$ changes the spread and shape of the posterior, but not the point estimate, which is therefore not robust to outliers.
Third, our asymptotic results assume a fixed number of covariates $p$ and a positive definite $\bm\Omega$; when $p$ is large relative to $n$, the construction still applies, since $\bm V_0$ is positive definite and $\bm\beta_n$ is a ridge-type shrinkage estimator, and with $\sigma_n^2=\sigma^2/a_n$ the posterior covariance $\sigma^2\bm V_n$ is exact also when $n<p$, but regimes in which $p$ grows with $n$ are not covered by our theory.
Fourth, the hyperparameters selected by the \textsc{loo-pll} are fixed in predictive resampling, so their uncertainty is not propagated, as shown by the equal widths for $f(\bm x)$ and $Q_{0.90}(\bm x)$ in Table~\ref{tab:sim}; a prior on them, as in Section~\ref{sec:loo_pll}, would address this.
Finally, the predictive scales are required to converge to a constant $\sigma$, so that the limiting model is homoscedastic; allowing a covariate-dependent limit would accommodate heteroscedastic responses, and we leave this extension to future work.

While the proposed method represents a substantial first step towards predictive Bayes methods for regression, an interesting direction for future work is to develop updates for a general nonparametric $f_n(\bm x)$ that does not assume linearity or additivity. In this direction, \citet{ng2025tabmgp, fortini2026principled} have recently provided a predictive Bayes interpretation of Tab\textsc{pfn}, a popular machine learning model for tabular data.
Although several candidates appear to yield convergent predictive rules, how to establish the martingale condition, or asymptotic exchangeability, for a general nonparametric estimator remains unclear. 
Predictive Bayes methods for multivariate responses, allowing inference on their dependence structure, are a further ambitious direction. 
More broadly, establishing theoretical guarantees for more general and flexible updates that retain computational scalability is an important next step.

\section*{Data availability}

The \textsc{nhanes} data analyzed in Section~\ref{sec:application} are publicly available from the website of the US National Center for Health Statistics (\texttt{www.cdc.gov/nchs/nhanes}). Code to download the data and to reproduce all the analyses is available in the repository \texttt{github.com/davideagno/BayesianRLS}.

\section*{Acknowledgments}

This research was partially supported by the National Institutes of Health (grant ID R01ES035625), by the Office of Naval Research (N000142412626), and by NSF award \#2426762.
The authors used Claude (Anthropic), an AI assistant, to help with language editing, with checking the consistency of the text and proofs, and with revising and rerunning the code. The authors reviewed all changes and take full responsibility for the content of the paper.

\section*{Appendix}
The Appendix contains the proofs of all theorems and propositions, additional simulation studies, additional results for the analysis of Section~\ref{sec:application}, and details on the computations. 
Code to replicate our analyses is available as described in the Data availability section.

% ============================================================
% Supplementary material (appendix)
% ============================================================
\clearpage
\appendix

\counterwithin{equation}{section}

\section{Proofs}
\label{apx:proofs}

This section collects the proofs of the results in Section~3 of the paper. We first prove three preliminary lemmas on the matrices $\bm V_n$, and then prove the results in the order in which they appear in the paper. Equation numbers without a letter prefix refer to the paper.

\subsection{Preliminary results}

\begin{lemma}
\label{lem:preceq}
For every $\bm u\in\mathds{R}^p$ and any $n\ge1$, $\bm u^\top\bm V_{n}\bm u\le\bm u^\top\bm V_{n-1}\bm u$.
\end{lemma}
\begin{proof}
The lemma states that the matrices $\bm V_n$ can only decrease as new observations arrive.
By Equation~(4), we can write $\bm V_{n-1}-\bm V_n=a_n\bm V_{n-1}\bm x_n\bm x_n^\top\bm V_{n-1}$, and therefore
\begin{equation*}
    \bm u^\top(\bm V_{n-1}-\bm V_n)\bm u
    = a_n\bm u^\top\bm V_{n-1}\bm x_n\bm x_n^\top\bm V_{n-1}\bm u
    = a_n(\bm u^\top\bm V_{n-1}\bm x_n)^2
    \ge 0,
\end{equation*}
since $a_n>0$. Rearranging the terms, $\bm u^\top\bm V_{n}\bm u\le\bm u^\top\bm V_{n-1}\bm u$.
\end{proof}

\begin{lemma}
\label{lem:xVx_to_zero}
Assume that the sequence of observed covariates $\bm x_{1:\infty}$ is known and assume that $\bm\Omega := \lim_{n\to\infty}n^{-1}\sum_{i=1}^n\bm x_i\bm x_i^\top$ exists and is positive definite.
Then, $\bm x_n^\top\bm V_{n-1}\bm x_n\to0$ as~$n\to\infty$.
\end{lemma}
\begin{proof}
We show that the largest eigenvalue of $\bm V_n$ decreases at rate $n^{-1}$, whereas $||\bm x_n||^2$ grows more slowly than $n$.
Since $\bm V_{n}$ is positive definite, for any $\bm u\in\mathds{R}^p$, $\bm u^\top\bm V_{n}\bm u\le\lambda_{\max}(\bm V_n)||\bm u||^2$, where $\lambda_{\max}(\bm V_n)$ denotes the maximum eigenvalue of the matrix $\bm V_n$.
Let $\lambda_* = \zeta\lambda_{\min}(\bm\Omega)$, with $\zeta\in(0,1)$.
Since by assumption $n^{-1}\sum_{i=1}^n\bm x_i\bm x_i^\top\to\bm\Omega$ as $n\to\infty$, with $\bm\Omega$ positive definite, there exists $n_0$ such that, for all $n\ge n_0$ and all $\bm u\in\mathds{R}^p$,
\begin{equation*}
    \bm u^\top\bigg( n^{-1}\sum_{i=1}^n\bm x_i\bm x_i^\top\bigg)\bm u \ge \bm u^\top\lambda_*\bm I_p\bm u,
\end{equation*}
and therefore, multiplying by $n$ both sides,
\begin{equation*}
     \bm u^\top\bigg(\sum_{i=1}^n\bm x_i\bm x_i^\top\bigg)\bm u \ge \bm u^\top n\lambda_*\bm I_p\bm u.
\end{equation*}
Then, by Equation~(5) and since $\bm u^\top\bm V_0^{-1}\bm u > 0$,
\begin{equation*}
    \bm u^\top\bm V_n^{-1}\bm u
    = \bm u^\top\bm V_0^{-1}\bm u + \bm u^\top\bigg( \sum_{i=1}^n\bm x_i\bm x_i^\top \bigg) \bm u
    > \bm u^\top\bigg( \sum_{i=1}^n\bm x_i\bm x_i^\top \bigg) \bm u
    \ge \bm u^\top n\lambda_*\bm I_p\bm u,
\end{equation*}
that is, all the eigenvalues of $\bm V_n^{-1}$ are larger than $n\lambda_*$. Therefore, for the inverse matrix $\bm V_n$, whose eigenvalues are the reciprocals of those of $\bm V_n^{-1}$, all the eigenvalues are smaller than $n^{-1}\lambda_*^{-1}$, that is, $\bm u^\top\bm V_n\bm u\le \bm u^\top n^{-1}\lambda_*^{-1}\bm I_p\bm u$ for all $\bm u\in\mathds R^p$, and $\lambda_{\max}(\bm V_n)\le n^{-1}\lambda_*^{-1}$.
Then, as~$n\to\infty$,
\begin{equation*}
0 \le \bm x_{n}^\top\bm V_{n-1}\bm x_n
\le \lambda_{\max}(\bm V_{n-1})||\bm x_{n}||^2
\le \lambda_*^{-1}(n-1)^{-1}||\bm x_n||^2 \longrightarrow 0,
\end{equation*}
since $n^{-1}||\bm x_n||^2\to0$. This last fact follows from
\begin{equation*}
    n^{-1}\bm x_n\bm x_n^\top=n^{-1}\sum_{i=1}^n\bm x_i\bm x_i^\top-\frac{n-1}{n}\bigg\{(n-1)^{-1}\sum_{i=1}^{n-1}\bm x_i\bm x_i^\top\bigg\}\longrightarrow\bm\Omega-\bm\Omega=\bm 0,
\end{equation*}
whose trace is $n^{-1}||\bm x_n||^2$.
\end{proof}

\begin{lemma}
\label{lem:sandwich}
Under Assumption~3, let $\bm\Sigma_n$ be as in Proposition~2 of the paper, and let $\epsilon_n:=\sup_{i>n}\bm x_i^\top\bm V_{i-1}\bm x_i$, $s_n^-:=\inf_{i>n}\sigma_i^2$ and $s_n^+:=\sup_{i>n}\sigma_i^2$. Then, for every $n\ge0$, $\bm\Sigma_n$ is positive definite and
\begin{equation*}
    s_n^-(1-\epsilon_n)\bm V_n\preceq\bm\Sigma_n\preceq s_n^+\bm V_n.
\end{equation*}
Here, for symmetric matrices $\bm A$ and $\bm B$, $\bm A\preceq\bm B$ means that $\bm u^\top\bm A\bm u\le\bm u^\top\bm B\bm u$
for all $\bm u\in\mathds{R}^p$, that is, $\bm B-\bm A$ is positive semidefinite.
\end{lemma}
\begin{proof}
Recall that $\bm\Sigma_n=\sum_{i>n}\sigma_i^2a_i(\bm V_{i-1}-\bm V_i)$, where $a_i=(1+\bm x_i^\top\bm V_{i-1}\bm x_i)^{-1}$. We show below that also $\bm V_n=\sum_{i>n}(\bm V_{i-1}-\bm V_i)$. Hence, $\bm\Sigma_n$ and $\bm V_n$ are weighted sums of the same positive semidefinite matrices $\bm V_{i-1}-\bm V_i$, with weights $\sigma_i^2a_i$ and $1$, respectively. The two bounds then follow by bounding the weights $\sigma_i^2a_i$, for $i>n$, from above by $s_n^+$ and from below by $s_n^-(1-\epsilon_n)$.
We first show that $(1-\epsilon_n)\bm V_n\preceq\bm S_n\preceq\bm V_n$, where $\bm S_n := \sum_{i>n}a_i(\bm V_{i-1}-\bm V_i)$ is the matrix $\bm\Sigma_n$ with all $\sigma_i^2$ replaced by $1$.
By Equation~(5) and Assumption~3, $n^{-1}\bm V_n^{-1} = n^{-1}\bm V_0^{-1} + n^{-1}\sum_{i=1}^n\bm x_i\bm x_i^\top\to\bm\Omega$, so that $\lim_{n\to\infty}\bm V_n=\bm0$.
Since the sum $\sum_{i=n+1}^N(\bm V_{i-1}-\bm V_i)=\bm V_n-\bm V_N$ telescopes, we can write
\begin{equation*}
    \bm V_n = \bm V_n-\lim_{N\to\infty}\bm V_N = \lim_{N\to\infty}\sum_{i=n+1}^N (\bm V_{i-1}-\bm V_i)
    = \sum_{i>n} (\bm V_{i-1}-\bm V_i).
\end{equation*}
The series defining $\bm S_n$ and $\bm\Sigma_n$ also converge, since their terms are positive semidefinite by Lemma~\ref{lem:preceq} and bounded, in the order $\preceq$, by $\bm V_{i-1}-\bm V_i$ and $\sigma_*^2(\bm V_{i-1}-\bm V_i)$, respectively, because $a_i\le1$ and $\sigma_i^2\le\sigma_*^2$.
Then,
\begin{equation*}
    \bm V_n-\bm S_n = \sum_{i>n}(\bm V_{i-1}-\bm V_i)-\sum_{i>n}a_i(\bm V_{i-1}-\bm V_i)
    = \sum_{i>n}(1-a_i)(\bm V_{i-1}-\bm V_i)
    \succeq \bm 0,
\end{equation*}
because each $\bm V_{i-1}-\bm V_i$ is positive semidefinite by Lemma~\ref{lem:preceq} and $a_i\le1$. This shows that $\bm S_n\preceq\bm V_n$. Moreover, for every~$i>n$,
\begin{equation*}
    0\le1-a_i=\frac{\bm x_i^\top\bm V_{i-1}\bm x_i}{1+\bm x_i^\top\bm V_{i-1}\bm x_i}\le \bm x_i^\top\bm V_{i-1}\bm x_i\le \epsilon_n,
\end{equation*}
where the second inequality holds because the denominator $a_i^{-1}=1+\bm x_i^\top\bm V_{i-1}\bm x_i$ is at least $1$, and the third because $\epsilon_n$ is the supremum of $\bm x_i^\top\bm V_{i-1}\bm x_i$ over $i>n$. Hence, recalling that $\bm V_n-\bm S_n=\sum_{i>n}(1-a_i)(\bm V_{i-1}-\bm V_i)$, and since each $\bm V_{i-1}-\bm V_i$ is positive semidefinite, so that the scalar bound $1-a_i\le\epsilon_n$ carries over to each term of the sum, we obtain
\begin{equation*}
    \bm V_n-\bm S_n\preceq\epsilon_n\sum_{i>n}(\bm V_{i-1}-\bm V_i)=\epsilon_n\bm V_n,
\end{equation*}
that is, $(1-\epsilon_n)\bm V_n\preceq\bm S_n$. Together with $\bm S_n\preceq\bm V_n$, shown above, this gives
\begin{equation*}
    (1-\epsilon_n)\bm V_n\preceq\bm S_n\preceq\bm V_n.
\end{equation*}

We now bring the weights $\sigma_i^2$ into play. By Lemma~\ref{lem:preceq}, $a_i(\bm V_{i-1}-\bm V_i)$ is positive semidefinite for every $i>n$. Then
\begin{equation*}
    \bm\Sigma_n-s_n^-\bm S_n = \sum_{i>n}(\sigma_i^2-s_n^-)a_i(\bm V_{i-1}-\bm V_i)\succeq \bm 0
\end{equation*}
since $\sigma_i^2-s_n^-\ge0$ for $i>n$, and analogously
\begin{equation*}
    s_n^+\bm S_n-\bm\Sigma_n = \sum_{i>n}(s_n^+ -\sigma_i^2)a_i(\bm V_{i-1}-\bm V_i)\succeq \bm 0
\end{equation*}
since $s_n^+-\sigma_i^2\ge0$ for $i>n$.
Therefore, $s_n^-\bm S_n\preceq\bm \Sigma_n\preceq s_n^+\bm S_n$ and, by what we have previously proved,
\begin{equation*}
    s_n^-(1-\epsilon_n)\bm V_n\preceq s_n^-\bm S_n\preceq\bm\Sigma_n\preceq s_n^+\bm S_n\preceq s_n^+\bm V_n.
\end{equation*}
Note also that $\epsilon_n\to0$ as $n\to\infty$: by Lemma~\ref{lem:xVx_to_zero}, $\bm x_i^\top\bm V_{i-1}\bm x_i\to0$ as $i\to\infty$, and $\epsilon_n$ is the supremum of the tail $(\bm x_i^\top\bm V_{i-1}\bm x_i)_{i>n}$ of this sequence. Hence, the lower bound is informative, that is, $1-\epsilon_n>0$, for all $n$ large enough.

Finally, $\bm\Sigma_n$ is positive definite for every $n\ge0$. For $i>n$, $a_i=(1+\bm x_i^\top\bm V_{i-1}\bm x_i)^{-1}\ge(1+\epsilon_n)^{-1}$ and $\sigma_i^2\ge s_n^-$, so that, since each $\bm V_{i-1}-\bm V_i$ is positive semidefinite,
\begin{equation*}
    \bm\Sigma_n=\sum_{i>n}\sigma_i^2a_i(\bm V_{i-1}-\bm V_i)\succeq\frac{s_n^-}{1+\epsilon_n}\sum_{i>n}(\bm V_{i-1}-\bm V_i)=\frac{s_n^-}{1+\epsilon_n}\bm V_n.
\end{equation*}
The right-hand side is positive definite, because $\bm V_n$ is positive definite, $\epsilon_n$ is finite, as the supremum of a sequence that converges to $0$, and $s_n^->0$, as the infimum of positive values that converge to $\sigma^2>0$ by Assumption~3.
\end{proof}

\subsection{Proof of Theorem 1}
\label{apx:th_1}
The proof follows the standard route for parametric martingale posteriors \citep[see, e.g.,][]{holmes2023statistical, fong2024asymptotics}. In part $(a)$, we show that, for every fixed $\bm u\in\mathds R^p$, the scalar sequence $\bm u^\top\bm\beta_n$ is a martingale bounded in $L_2$, and conclude with Doob's martingale convergence theorem \citep{doob1953stochastic}. In part $(b)$, we derive asymptotic exchangeability from the almost sure weak convergence of the one-step predictive distributions, a standard predictive approach to exchangeability \citep{fortini2012predictive, fortini2024exchangeability}.

\begin{itemize}
\item[$(a)$] Let $\bm u\in\mathds{R}^p$ and $\theta_n = \theta_n(\bm u) = \bm u^\top\bm\beta_n$. Multiplying Equation~(3) on the left by $\bm u^\top$ gives
\begin{equation*}
    \bm u^\top\bm\beta_n=\bm u^\top\bm\beta_{n-1}+a_n\bm u^\top\bm V_{n-1}\bm X_n(Y_n-\bm X_n^\top\bm\beta_{n-1}),
    \qquad\text{that is,}\qquad
    \theta_n=\theta_{n-1} + c_n e_n,
\end{equation*}
where $c_n := a_n\bm u^\top\bm V_{n-1}\bm X_n$ and $e_n:=Y_n-\bm X_n^\top\bm\beta_{n-1}$.
Let $(\mathcal{F}_n)_{n\ge0}$ be the filtration with $\mathcal{F}_0$ the trivial $\sigma$-field and $\mathcal{F}_n=\sigma(\bm X_1,Y_1,\ldots,\bm X_n,Y_n)$ for $n\ge1$; note that $\theta_0=\bm u^\top\bm\beta_0$ is a constant. Both $c_n$ and $\sigma_n=s_{n-1}(\bm X_n)$ are measurable with respect to $\mathcal F_{n-1}\vee\sigma(\bm X_n)$, that is, they are known once the past observations and the new covariate $\bm X_n$ are known.
By Equation~(2) and Assumption~1, given $\mathcal F_{n-1}$ and $\bm X_n$, the response $Y_n$ has mean $\bm X_n^\top\bm\beta_{n-1}$ and variance $\sigma_n^2$, so that
\begin{equation*}
    \mathds{E}(e_n\mid\mathcal{F}_{n-1},\bm X_n) = 0,
    \qquad
    \mathds{E}(e_n^2\mid\mathcal{F}_{n-1},\bm X_n)=\sigma_n^2\le\sigma_*^2.
\end{equation*}
Consequently, by the tower property of conditional expectations,
\begin{align*}
    \mathds{E}(\theta_n\mid\mathcal{F}_{n-1})
    & = \mathds{E}(\theta_{n-1}+c_n e_n\mid\mathcal{F}_{n-1})\\
    & = \theta_{n-1}+\mathds{E}\{c_n\,\mathds{E}(e_n\mid\mathcal{F}_{n-1},\bm X_n)\mid \mathcal{F}_{n-1}\}
    = \theta_{n-1}.
\end{align*}
Then, $(\theta_n)_{n\ge0}$ is a martingale with respect to the filtration $(\mathcal{F}_n)_{n\ge0}$; integrability, $\mathds E|\theta_n|<\infty$, follows from the bound on $\mathds{E}(\theta_n^2)$ below.

Now we show that $\sup_{n}\mathds{E}(\theta_n^2)<\infty$.
Squaring $\theta_n=\theta_{n-1}+c_ne_n$ and conditioning on $\mathcal F_{n-1}$ and $\bm X_n$, with respect to which $\theta_{n-1}$ and $c_n$ are measurable, we obtain, from the two conditional moments of $e_n$ displayed above,
\begin{align*}
    \mathds{E}(\theta_n^2\mid\mathcal{F}_{n-1},\bm X_n)
    & = \theta_{n-1}^2 + 2\theta_{n-1}c_n\,\mathds{E}(e_n\mid\mathcal{F}_{n-1},\bm X_n) + c_n^2\,\mathds{E}(e_n^2\mid\mathcal{F}_{n-1},\bm X_n)\\
    & = \theta_{n-1}^2 + \sigma_n^2c_n^2 \le \theta_{n-1}^2 + \sigma_*^2c_n^2.
\end{align*}
Taking expectations and iterating over $n$ gives $\mathds{E}(\theta_n^2)\le\theta_0^2+\sigma_*^2\,\mathds{E}(\sum_{i=1}^n c_i^2)$.
The quantity $c_n^2$ can be rewritten using Equation~(4), which gives $\bm V_{n-1}-\bm V_n=a_n\bm V_{n-1}\bm X_n\bm X_n^\top\bm V_{n-1}$:
\begin{equation*}
    c_n^2 = a_n^2(\bm u^\top\bm V_{n-1}\bm X_n)^2
    = a_n\bm u^\top(\bm V_{n-1}-\bm V_n)\bm u \le \bm u^\top\bm V_{n-1}\bm u - \bm u^\top\bm V_n\bm u,
\end{equation*}
since $a_n\in(0,1]$ and $\bm u^\top(\bm V_{n-1}-\bm V_n)\bm u\ge0$ by Lemma~\ref{lem:preceq}.
Summing these inequalities over $i=1,\ldots,n$, the right-hand side telescopes, so that, for every realization of the covariates,
\begin{equation*}
    \sum_{i=1}^n c_i^2 \le \sum_{i=1}^n\big(\bm u^\top\bm V_{i-1}\bm u - \bm u^\top\bm V_i\bm u\big)
    = \bm u^\top\bm V_0\bm u - \bm u^\top\bm V_n\bm u
    \le \bm u^\top\bm V_0\bm u,
\end{equation*}
where the last inequality holds because $\bm V_n$ is positive definite. Hence, for every $n\ge1$,
\begin{equation*}
    \mathds{E}(\theta_n^2)\le \theta_0^2 + \sigma_*^2\,\bm u^\top\bm V_0\bm u,
\end{equation*}
which is finite, since $\theta_0=\bm u^\top\bm\beta_0$ is fixed, and does not depend on~$n$.
Therefore, $(\theta_n)_{n\ge0}$ is a martingale bounded in $L_2$. By Doob's convergence theorem for such martingales \citep[Theorem~12.1]{williams1991probability}, $\theta_n$ converges almost surely and in $L_2$ to a limit random variable $\tilde\theta(\bm u)$. Moreover, since convergence in $L_2$ implies convergence in $L_1$, and conditional expectations are continuous in $L_1$, letting $m\to\infty$ in $\mathds{E}(\theta_m\mid\mathcal F_n)=\theta_n$, for $m>n$, gives $\mathds{E}\{\tilde\theta(\bm u)\mid\mathcal F_n\}=\theta_n$.
Taking $\bm u$ equal to each vector of the canonical basis of $\mathds R^p$, we obtain a random vector $\tilde{\bm\beta}$ such that $\bm\beta_n$ converges to $\tilde{\bm\beta}$ almost surely and in $L_2$, and $\mathds{E}(\tilde{\bm\beta}\mid\mathcal{F}_n)=\bm\beta_n$.

\item[$(b)$] By Lemma~8.2(b) in \cite{aldous1983exchangeability}, a sequence of random elements of a Polish space is asymptotically exchangeable if its one-step predictive distributions converge weakly almost surely to a random probability measure, which is then its directing random measure. We identify the predictive distribution~$Q_n$ of $(\bm X_{n+1},Y_{n+1})$ in Step~1 and prove its convergence in Step~2; since $Q_n$ depends on~$n$ only through $\bm\beta_n$ and~$s_n$, this follows from part~$(a)$ and Assumption~2. Step~3 applies the lemma.

\medskip
\emph{Step 1: the predictive distribution.}
Let $\bm R_n=(\bm X_n,Y_n)\in S:=\mathds R^p\times\mathds R$, so that $\mathcal F_n=\sigma(\bm R_{1:n})$. By Equation~(2) and Assumptions~1--2, $\bm R_{n+1}=(\bm X_{n+1},\bm X_{n+1}^\top\bm\beta_n+s_n(\bm X_{n+1})\xi_{n+1})$, where $\xi_{n+1}$ is defined by this identity, since $s_n>0$, and $\bm X_{n+1}\sim P_X$ and $\xi_{n+1}\sim P_\xi$ are independent of each other and of $\mathcal F_n$: indeed, $\bm X_{n+1}$ is independent of $\mathcal F_n$ by Assumption~2, and the conditional law of $\xi_{n+1}$ given $\mathcal F_n$ and $\bm X_{n+1}$ is $P_\xi$, which does not depend on them. Hence, holding $\bm R_{1:n}$ fixed and integrating over $(\bm X_{n+1},\xi_{n+1})$ \citep[Exercise~IV.2.26]{cinlar2011probability}, a regular version of $\mathds P(\bm R_{n+1}\in\cdot\mid\mathcal F_n)$ is
\begin{equation*}
    Q_n(\mathrm d\bm x,\mathrm dy)=P_X(\mathrm d\bm x)\,P\{\mathrm dy\mid\bm x,\bm\beta_n,s_n(\bm x)\}.
\end{equation*}

\medskip
\emph{Step 2: weak convergence of the predictive distributions.}
Let $\Omega_0$ be the set on which $\bm\beta_n\to\tilde{\bm\beta}$ and $s_n(\bm x)\to\sigma$ for $P_X$-almost every $\bm x$. By Fubini's theorem, $\Omega_0$ is an event, since $s_n$ is jointly measurable in $\bm x$ and $\bm X_{1:n}$, and it has probability one by part~$(a)$ and Assumption~2. Fix $\omega\in\Omega_0$. For $P_X$-almost every $\bm x$, we have $\bm x^\top\bm\beta_n+s_n(\bm x)\xi\to\bm x^\top\tilde{\bm\beta}+\sigma\xi$ for every $\xi\in\mathds R$; taking $\xi\sim P_\xi$, by Assumption~1, $P\{\cdot\mid\bm x,\bm\beta_n,s_n(\bm x)\}\to P(\cdot\mid\bm x,\tilde{\bm\beta},\sigma)$ weakly, as pointwise convergence implies convergence in distribution. Hence, for every bounded continuous $f:S\to\mathds R$, since $y\mapsto f(\bm x,y)$ is bounded and continuous, the inner integral $\int f(\bm x,y)\,P\{\mathrm dy\mid\bm x,\bm\beta_n,s_n(\bm x)\}$ converges to $\int f(\bm x,y)\,P(\mathrm dy\mid\bm x,\tilde{\bm\beta},\sigma)$ for $P_X$-almost every $\bm x$, and it is bounded by $\|f\|_\infty:=\sup|f|$. Integrating with respect to $P_X$, dominated convergence gives $\int f\,\mathrm dQ_n\to\int f\,\mathrm d\tilde P_{\bm X,Y}$, where
\begin{equation*}
    \tilde P_{\bm X,Y}(\mathrm d\bm x,\mathrm dy)=P_X(\mathrm d\bm x)\,P(\mathrm dy\mid\bm x,\tilde{\bm\beta},\sigma)
\end{equation*}
is a random probability measure, by the measurability of $\tilde{\bm\beta}$ and Fubini's theorem. Thus, $Q_n\to\tilde P_{\bm X,Y}$ weakly on $\Omega_0$, that is, $\mathds P$-almost surely.

\medskip
\emph{Step 3: asymptotic exchangeability.}
Since $S$ is Polish, by Lemma~8.2(b) in \cite{aldous1983exchangeability} the almost sure weak convergence of the one-step predictive distributions implies that $(\bm R_n)_{n\ge1}$ is asymptotically exchangeable with directing random measure $\tilde P_{\bm X,Y}$.
By construction, its $\bm X$-marginal is $P_X$ and $P(\cdot\mid\bm x,\tilde{\bm\beta},\sigma)$ is a version of its conditional distribution of $Y$ given~$\bm X=\bm x$.
\end{itemize}

\subsection{Proof of Proposition 1}
The proof parallels that of Theorem~1. Part $(a)$ repeats the martingale argument, now with known covariates. In part $(b)$, the predictive distribution $P_n$ need not converge, because $\bm x_{n+1}$ changes with $n$; we show instead, by a coupling argument, that it becomes close to $P(\cdot\mid\bm x_{n+1},\tilde{\bm\beta},\sigma)$, in the sense of weak merging \citep{daristotile1988merging}, a notion weaker than the merging in total variation of \citet{blackwell1962merging}. We conclude by deriving the series representation of $\tilde{\bm\beta}$ in Equation~(6) of the paper.

\begin{itemize}
\item[$(a)$] The argument is that of part $(a)$ of Theorem~1, with three simplifications. Under Assumption~3 the covariates are known, so that $c_n=a_n\bm u^\top\bm V_{n-1}\bm x_n$ and $\sigma_n$ are deterministic, and $e_n=Y_n-\bm x_n^\top\bm\beta_{n-1}$; the filtration is $\mathcal F_n=\sigma(Y_1,\ldots,Y_n)$ for $n\ge1$, with $\mathcal F_0$ the trivial $\sigma$-field; and it suffices to condition on $\mathcal F_{n-1}$, rather than on $\mathcal F_{n-1}$ and $\bm X_n$. Indeed, by Equation~(2) and Assumption~1, $\mathds{E}(e_n\mid\mathcal{F}_{n-1})=0$ and $\mathds{E}(e_n^2\mid\mathcal{F}_{n-1})=\sigma_n^2\le\sigma_*^2$, so that the same computations show that $(\theta_n)_{n\ge0}$ is a martingale with $\mathds{E}(\theta_n^2)\le \theta_0^2 + \sigma_*^2\bm u^\top\bm V_0\bm u$ for every $n$. Doob's martingale convergence theorem then gives, as in Theorem~1, that $\bm\beta_n\to\tilde{\bm\beta}$ almost surely and in $L_2$, and that $\mathds E(\tilde{\bm\beta}\mid\mathcal F_n)=\bm\beta_n$.

\item[$(b)$] Fix an outcome on which $\bm\beta_n\to\tilde{\bm\beta}$, which has $\mathds{P}_{\bm x_{1:\infty}}$-probability one by part $(a)$; on this outcome, $\bm\beta_n$ and $\tilde{\bm\beta}$ are fixed vectors. Let $\xi\sim P_\xi$ be an auxiliary random variable. By Assumption~1, the random variables
\begin{equation*}
    Y'_{n+1}:=\bm x_{n+1}^\top\bm\beta_n+\sigma_{n+1}\xi
    \qquad\text{and}\qquad
    \tilde Y_{n+1}:=\bm x_{n+1}^\top\tilde{\bm\beta}+\sigma\xi
\end{equation*}
have distributions $P_n$ and $P(\cdot\mid\bm x_{n+1},\tilde{\bm\beta},\sigma)$, respectively. Thus, $Y'_{n+1}$ has the same law as $Y_{n+1}$ given $y_{1:n}$, but, unlike $Y_{n+1}$, it is built from the same $\xi$ as $\tilde Y_{n+1}$.
Let $f:\mathds R\to\mathds R$ be bounded and continuous, and let
\begin{equation*}
    d_n:=\int f\,\mathrm dP_n-\int f\,\mathrm dP(\cdot\mid\bm x_{n+1},\tilde{\bm\beta},\sigma)=\mathds E\{f(Y'_{n+1})-f(\tilde Y_{n+1})\}.
\end{equation*}
We need to show that $d_n\to0$ as $n\to\infty$, for every such~$f$. Unlike in the proof of Theorem~1$(b)$, we cannot let $n\to\infty$ in each of the two terms, because $\bm x_{n+1}$, and hence $Y'_{n+1}$ and $\tilde Y_{n+1}$, need not converge. We will show that $Y'_{n+1}-\tilde Y_{n+1}\to0$, but this alone does not imply that $f(Y'_{n+1})-f(\tilde Y_{n+1})\to0$, because a bounded continuous $f$ need not be uniformly continuous on~$\mathds R$, as for $f(y)=\sin(y^2)$. Therefore, we also show that both variables lie in a compact interval, independent of~$n$, on which $f$ is uniformly continuous.

Let $M:=\sup_{n\ge1}||\bm x_n||$, which is finite because the sequence $\bm x_{1:\infty}$ is bounded by Assumption~3. Subtracting the two definitions, $Y'_{n+1}-\tilde Y_{n+1}=\bm x_{n+1}^\top(\bm\beta_n-\tilde{\bm\beta})+(\sigma_{n+1}-\sigma)\xi$, and therefore
\begin{align*}
    |Y'_{n+1}-\tilde Y_{n+1}|
    & \le|\bm x_{n+1}^\top(\bm\beta_n-\tilde{\bm\beta})|+|\sigma_{n+1}-\sigma|\,|\xi|\\
    & \le||\bm x_{n+1}||\,||\bm\beta_n-\tilde{\bm\beta}||+|\sigma_{n+1}-\sigma|\,|\xi|\\
    & \le M||\bm\beta_n-\tilde{\bm\beta}||+|\sigma_{n+1}-\sigma|\,|\xi|,
\end{align*}
where the first inequality is the triangle inequality, the second is the Cauchy--Schwarz inequality applied to the vectors $\bm x_{n+1}$ and $\bm\beta_n-\tilde{\bm\beta}$, and the third holds because $||\bm x_{n+1}||\le M$. The right-hand side tends to $0$, because $\bm\beta_n\to\tilde{\bm\beta}$ and $\sigma_n\to\sigma$ by Assumption~3, so that $Y'_{n+1}-\tilde Y_{n+1}\to0$ for every value of~$\xi$.

For the interval, note that on the fixed outcome $(\bm\beta_n)_{n\ge0}$ is a convergent sequence in $\mathds R^p$, hence bounded, and its limit satisfies $||\tilde{\bm\beta}||\le\sup_{n\ge0}||\bm\beta_n||$; moreover, $\sigma_{n+1}\le\sigma_*$ and $\sigma\le\sigma_*$, the latter because $\sigma$ is the limit of~$\sigma_n$. The same two inequalities as above then give $|Y'_{n+1}|\le M||\bm\beta_n||+\sigma_{n+1}|\xi|$ and $|\tilde Y_{n+1}|\le M||\tilde{\bm\beta}||+\sigma|\xi|$, so that both variables lie in the interval $[-C-\sigma_*|\xi|,\,C+\sigma_*|\xi|]$, with $C:=M\sup_{n\ge0}||\bm\beta_n||<\infty$, for every~$n$.

By the Heine--Cantor theorem, $f$ is uniformly continuous on this compact interval. Since $Y'_{n+1}-\tilde Y_{n+1}\to0$ and both variables lie in it, it follows that $f(Y'_{n+1})-f(\tilde Y_{n+1})\to0$ for every value of~$\xi$. As $|f(Y'_{n+1})-f(\tilde Y_{n+1})|\le2\|f\|_\infty$, dominated convergence with respect to~$P_\xi$ gives $d_n\to0$.
Since the outcome was arbitrary in an event of probability one, $P_n$ and $P(\cdot\mid\bm x_{n+1},\tilde{\bm\beta},\sigma)$ merge weakly $\mathds{P}_{\bm x_{1:\infty}}$-almost surely.
\end{itemize}

\medskip
Finally, we derive Equation~(6) of the paper. For $i\ge1$, let $\xi_i:=(Y_i-\bm x_i^\top\bm\beta_{i-1})/\sigma_i$. By Equation~(2) and Assumption~1, the $\xi_i$ are iid from $P_\xi$, and those with $i>n$ are independent of~$y_{1:n}$.
By Equation~(3), $\bm\beta_i-\bm\beta_{i-1}=\bm w_i\xi_i$, with $\bm w_i=\sigma_ia_i\bm V_{i-1}\bm x_i$ deterministic under Assumption~3. Summing from $i=n+1$ to $N$, the left-hand side telescopes, so that $\bm\beta_N-\bm\beta_n=\sum_{i=n+1}^N\bm w_i\xi_i$. As $N\to\infty$, $\bm\beta_N\to\tilde{\bm\beta}$ almost surely and in $L_2$ by part~$(a)$, and hence the partial sums converge, in the same senses, to $\tilde{\bm\beta}-\bm\beta_n$. This gives
\begin{equation*}
    \tilde{\bm\beta}-\bm\beta_n=\sum_{i>n}\bm w_i\xi_i,
\end{equation*}
where the series converges almost surely and in $L_2$. In particular, $\tilde{\bm\beta}-\bm\beta_n$ depends only on $\xi_{n+1},\xi_{n+2},\ldots$ and is therefore independent of~$y_{1:n}$.

\subsection{Proof of Proposition 2}

By Equation~(6) of the paper, derived at the end of the proof of Proposition~1, $\tilde{\bm\beta}-\bm\beta_n=\sum_{i>n}\bm w_i\xi_i$ is independent of $y_{1:n}$, with deterministic weights and $\xi_i$ iid with zero mean and unit variance. Hence, its partial sums $\sum_{i=n+1}^N\bm w_i\xi_i$ have variance $\sum_{i=n+1}^N\bm w_i\bm w_i^\top=\bm\Sigma_n-\bm\Sigma_N$. They converge in $L_2$ as $N\to\infty$, so that their variances converge, by the Cauchy--Schwarz inequality, and $\bm\Sigma_N\to\bm0$, as the tail of the series defining $\bm\Sigma_0$, which converges by the proof of Lemma~\ref{lem:sandwich}. Therefore, since $\bm\beta_n$ is a function of $y_{1:n}$,
\begin{equation*}
    \var(\tilde{\bm\beta}\mid y_{1:n})=\var(\tilde{\bm\beta}-\bm\beta_n\mid y_{1:n})=\var(\tilde{\bm\beta}-\bm\beta_n)=\lim_{N\to\infty}(\bm\Sigma_n-\bm\Sigma_N)=\bm\Sigma_n.
\end{equation*}
The same argument with $n=0$, where $\bm\beta_0$ is fixed, gives $\var(\tilde{\bm\beta})=\bm\Sigma_0$.
When $\sigma_n^2=\sigma^2/a_n$, the weights $\sigma_i^2a_i$ are all equal to $\sigma^2$, so that $\bm\Sigma_n=\sigma^2\sum_{i>n}(\bm V_{i-1}-\bm V_i)=\sigma^2\bm V_n$, where the last equality is shown in the proof of Lemma~\ref{lem:sandwich}.

\subsection{Proof of Proposition 3}
Both limits follow from Equation~(5) and the bounds of Lemma~\ref{lem:sandwich}. For $\bm V_n$, by Equation~(5), $n^{-1}\bm V_n^{-1} = n^{-1}\bm V_0^{-1} + n^{-1}\sum_{i=1}^n\bm x_i\bm x_i^\top\to\bm\Omega$, and therefore, by continuity of matrix inversion, $n\bm V_n\to\bm\Omega^{-1}$ as $n\to\infty$.

For $\bm\Sigma_n$, multiplying by $n$ the inequalities of Lemma~\ref{lem:sandwich} gives
\begin{equation*}
    s_n^-(1-\epsilon_n)\,n\bm V_n\preceq n\bm\Sigma_n\preceq s_n^+\,n\bm V_n.
\end{equation*}
As $n\to\infty$, $\epsilon_n\to0$ by Lemma~\ref{lem:xVx_to_zero}, $s_n^-\to\sigma^2$ and $s_n^+\to\sigma^2$, since they are the infimum and the supremum of the tail $(\sigma_i^2)_{i>n}$ of a sequence that converges to $\sigma^2$ by Assumption~3, and $n\bm V_n\to\bm\Omega^{-1}$. Hence both bounds converge to $\sigma^2\bm\Omega^{-1}$, so that $\bm u^\top n\bm\Sigma_n\bm u\to\sigma^2\bm u^\top\bm\Omega^{-1}\bm u$ for every $\bm u\in\mathds R^p$. Since the matrices are symmetric, this implies $n\bm\Sigma_n\to\sigma^2\bm\Omega^{-1}$.

\subsection{Proof of Theorem 2}

Asymptotic normality of martingale posteriors is typically derived from martingale central limit theorems \citep{fortini_petrone_2020, fong2024asymptotics}. Here, the terms of the series in Equation~(6) of the paper are independent, and we apply the Lindeberg--Feller theorem to the standardized version of $\tilde{\bm\beta}-\bm\beta_n$.

\medskip
\emph{Step 1: representation.}
By Equation~(6) of the paper, derived at the end of the proof of Proposition~1,
\begin{equation*}
  \tilde{\bm\beta}-\bm\beta_n=\sum_{i>n}\bm w_i\,\xi_i,
  \qquad\mathrm{and}\qquad
  \sum_{i>n}\bm w_i\bm w_i^\top=\sum_{i>n}\sigma_i^2a_i(\bm V_{i-1}-\bm V_i)=\bm\Sigma_n,
\end{equation*}
with deterministic weights $\bm w_i=\sigma_ia_i\bm V_{i-1}\bm x_i$, $\xi_i$ iid from $P_\xi$, and the series converging almost surely and in $L_2$. Since $\tilde{\bm\beta}-\bm\beta_n$ is independent of $y_{1:n}$, its conditional distribution given $y_{1:n}$ coincides with its distribution under $\mathds{P}_{\bm x_{1:\infty}}$, and it suffices to prove convergence in distribution.

\medskip
\emph{Step 2: reduction to the Lindeberg condition.}
Consider the standardized vector
\begin{equation*}
  \bm W_n:=\bm\Sigma_n^{-1/2}(\tilde{\bm\beta}-\bm\beta_n)=\sum_{i>n}\bm\Sigma_n^{-1/2}\bm w_i\,\xi_i=\sum_{i>n}\bm v_{n,i}\,\xi_i,
  \qquad \bm v_{n,i}:=\bm\Sigma_n^{-1/2}\bm w_i,
\end{equation*}
where $\bm\Sigma_n$ is invertible by Lemma~\ref{lem:sandwich} and the second equality follows from Equation~(6) of the paper. By Step~1, $\sum_{i>n}\bm v_{n,i}\bm v_{n,i}^\top=\bm\Sigma_n^{-1/2}\bm\Sigma_n\bm\Sigma_n^{-1/2}=\bm I_p$.
The Lindeberg--Feller theorem concerns sums of finitely many independent terms, whereas $\bm W_n$ is an infinite series. We therefore split $\bm W_n$ into a finite sum, to which the theorem applies, and a remainder, which we show to be negligible. Taking traces in the identity $\sum_{i>n}\bm v_{n,i}\bm v_{n,i}^\top=\bm I_p$, and recalling that $||\bm v||^2=\mathrm{tr}(\bm v\bm v^\top)$, gives $\sum_{i>n}||\bm v_{n,i}||^2=\mathrm{tr}(\bm I_p)=p<\infty$. As this series converges, we can choose $N_n>n$ such that $\sum_{i>N_n}||\bm v_{n,i}||^2\le n^{-1}$, and write $\bm W_n=\bm T_n+\bm U_n$, with
\begin{equation*}
  \bm T_n:=\sum_{i=n+1}^{N_n}\bm v_{n,i}\,\xi_i
  \qquad\mathrm{and}\qquad
  \bm U_n:=\sum_{i>N_n}\bm v_{n,i}\,\xi_i.
\end{equation*}
Any bound that tends to $0$ would do in place of $n^{-1}$: it makes the remainder vanish.

For the remainder, the terms $\bm v_{n,i}\xi_i$ have zero mean and are uncorrelated, because the $\xi_i$ are iid with zero mean and unit variance. Hence, the partial sums of $\bm U_n$ have expected squared norm equal to the corresponding partial sums of $||\bm v_{n,i}||^2$, and, passing to the limit by the $L_2$ convergence of the series in Step~1, $\mathds{E}(||\bm U_n||^2)=\sum_{i>N_n}||\bm v_{n,i}||^2\le n^{-1}$. This shows that $\bm U_n\to0$ in $L_2$, and therefore in probability, as $n\to\infty$.

For the finite part, $\bm T_n$ is a sum of the $N_n-n$ independent vectors $\bm v_{n,i}\xi_i$, $i=n+1,\ldots,N_n$, which have zero mean and covariance matrices $\bm v_{n,i}\bm v_{n,i}^\top$. These vectors form a triangular array, as they change with $n$. Since $\sum_{i>n}\bm v_{n,i}\bm v_{n,i}^\top=\bm I_p$, their total covariance is
\begin{equation*}
  \sum_{i=n+1}^{N_n}\bm v_{n,i}\bm v_{n,i}^\top=\bm I_p-\sum_{i>N_n}\bm v_{n,i}\bm v_{n,i}^\top\longrightarrow\bm I_p,
\end{equation*}
because the last sum is positive semidefinite with trace $\sum_{i>N_n}||\bm v_{n,i}||^2\le n^{-1}$, and each entry of a positive semidefinite matrix is bounded in absolute value by its trace. By the Lindeberg--Feller theorem \citep[Proposition~2.27]{van2000asymptotic}, if
\begin{equation}
\label{eqn:lindeberg}
\sum_{i=n+1}^{N_n}\mathds{E}\big\{||\bm v_{n,i}\xi_i||^2\mathds{1}(||\bm v_{n,i}\xi_i||>\epsilon)\big\}\to 0
\end{equation}
for every $\epsilon>0$ as $n\to\infty$, then $\bm T_n\to\bm Z$ in distribution, with $\bm Z\sim\mathcal N_p(\bm 0,\bm I_p)$. Together with $\bm U_n\to0$ in probability, Slutsky's theorem then gives $\bm W_n=\bm T_n+\bm U_n\to\bm Z$ in distribution.

\medskip
\emph{Step 3: the Lindeberg condition.}
Let $\rho_n:= \sup_{i>n}\bm w_i^\top\bm\Sigma_n^{-1}\bm w_i$, the largest weight of a single term.
Since $||\bm\Sigma_n^{-1/2}\bm w_i\xi_i||^2 =||\bm\Sigma_n^{-1/2}\bm w_i||^2\xi_i^2 =\bm w_i^\top\bm\Sigma_n^{-1}\bm w_i\xi_i^2$, and $||\bm\Sigma_n^{-1/2}\bm w_i\xi_i||\ge0$,
\begin{align*}
    \sum_{i=n+1}^{N_n}\mathds{E}\big\{||\bm v_{n,i}\xi_i||^2\mathds{1}(||\bm v_{n,i}\xi_i||>\epsilon)\big\} & \le \sum_{i>n}\mathds{E}\big\{||\bm\Sigma_n^{-1/2}\bm w_i\xi_i||^2\mathds{1}(||\bm\Sigma_n^{-1/2}\bm w_i\xi_i||>\epsilon)\big\}\\
    & = \sum_{i>n}\bm w_i^\top\bm\Sigma_n^{-1}\bm w_i \mathds{E}\big\{\xi_i^2\mathds{1}(\bm w_i^\top\bm\Sigma_n^{-1}\bm w_i\xi_i^2>\epsilon^2)\big\}\\
    & \le \sum_{i>n}\bm w_i^\top\bm\Sigma_n^{-1}\bm w_i \mathds{E}\big\{\xi_i^2\mathds{1}(\rho_n\xi_i^2>\epsilon^2)\big\}\\
    & = \sum_{i>n}\bm w_i^\top\bm\Sigma_n^{-1}\bm w_i \mathds{E}\big\{\xi_i^2\mathds{1}(\xi_i^2>\epsilon^2/\rho_n)\big\}\\
    & = p\, \mathds{E}\big\{\xi^2 \mathds{1}(\xi^2>\epsilon^2/\rho_n)\big\},
\end{align*}
because $\mathds{E}\big\{\xi_i^2\mathds{1}(\xi_i^2>\epsilon^2/\rho_n)\big\}$ is the same for every $i>n$ and
$\sum_{i>n}\bm w_i^\top\bm\Sigma_n^{-1}\bm w_i=\mathrm{tr}(\bm\Sigma_n^{-1}\bm\Sigma_n)=p$.
Since $\mathds E(\xi^2)=1<\infty$, the last expectation tends to $0$ by dominated convergence whenever $\epsilon^2/\rho_n\to\infty$. Therefore, \eqref{eqn:lindeberg} is satisfied if $\rho_n\to0$ as $n\to\infty$.

To show that $\rho_n\to0$, let $\epsilon_n$ and $s_n^-$ be as in Lemma~\ref{lem:sandwich}, and take $n$ large enough that $\epsilon_n<1$. By Lemma~\ref{lem:sandwich}, $\bm\Sigma_n\succeq s_n^-(1-\epsilon_n)\bm V_n$, where both sides are positive definite. Since inversion reverses the ordering between positive definite matrices, that is, $\bm A\preceq\bm B$ implies $\bm B^{-1}\preceq\bm A^{-1}$, we obtain $\bm\Sigma_n^{-1}\preceq \{s_n^-(1-\epsilon_n)\}^{-1}\bm V_n^{-1}$.
Moreover, for $i>n$, $\bm V_{i-1}\preceq\bm V_n$ by Lemma~\ref{lem:preceq}, and thus, for the inverses, $\bm V_n^{-1}\preceq\bm V_{i-1}^{-1}$. Multiplying on both sides by the symmetric matrix $\bm V_{i-1}$, which preserves the ordering since $\bm A\preceq\bm B$ implies $\bm C\bm A\bm C\preceq\bm C\bm B\bm C$ for every symmetric $\bm C$, we obtain $\bm V_{i-1}\bm V_n^{-1}\bm V_{i-1}\preceq\bm V_{i-1}\bm V_{i-1}^{-1}\bm V_{i-1}=\bm V_{i-1}$.
Therefore, since $a_i\le1$ and $\sigma_i\le\sigma_*$,
\begin{equation*}
  \bm w_i^\top\bm\Sigma_n^{-1}\bm w_i
  \le\frac{\sigma_*^2}{s_n^-(1-\epsilon_n)}\,
     \bm x_i^\top\bm V_{i-1}\bm V_n^{-1}\bm V_{i-1}\bm x_i
  \le\frac{\sigma_*^2}{s_n^-(1-\epsilon_n)}\,\bm x_i^\top\bm V_{i-1}\bm x_i
  \le\frac{\sigma_*^2}{s_n^-(1-\epsilon_n)}\,\epsilon_n,
\end{equation*}
so $\rho_n\to0$, because $\epsilon_n\to0$ by Lemma~\ref{lem:xVx_to_zero} and
$s_n^-\to\sigma^2>0$ by Assumption~3. This proves the first statement.

\medskip
\emph{Step 4: the second statement.}
We can write $\sigma^{-1}\bm V_n^{-1/2}(\tilde{\bm\beta}-\bm\beta_n)=\bm G_n\bm W_n$, with
\begin{equation*}
    \bm G_n:=\sigma^{-1}\bm V_n^{-1/2}\bm\Sigma_n^{1/2}=(\sigma^2n\bm V_n)^{-1/2}(n\bm\Sigma_n)^{1/2}\longrightarrow(\sigma^2\bm\Omega^{-1})^{-1/2}(\sigma^2\bm\Omega^{-1})^{1/2}=\bm I_p
\end{equation*}
by Proposition~3 and the continuity of the matrix square root. Since $\bm W_n\to\bm Z$ in distribution, Slutsky's theorem gives $\bm G_n\bm W_n\to\bm Z$ in distribution, which is the second statement.

\subsection{Proof of Theorem 3}

By Equation~(6) of the paper, $\tilde{\bm\beta} = \bm\beta_n + \sum_{i>n}\bm w_i \xi_i$, where $\xi_i\overset{\mathrm{iid}}{\sim}\mathcal{N}(0,1)$ in this case, so that $\bm w_i\xi_i\sim\mathcal{N}_p(\bm0,\bm w_i\bm w_i^\top)$.
Since $\bm w_i$ only depends on the known covariates, the vectors $\bm w_i\xi_i$, $i\ge1$, are independent, and those with $i>n$ are independent of $y_{1:n}$.
Each partial sum is a finite sum of independent Gaussian vectors, hence $\sum_{i=n+1}^N\bm w_i\xi_i\sim\mathcal N_p(\bm 0,\bm\Sigma_n-\bm\Sigma_N)$, with characteristic function $\exp\{-\bm t^\top(\bm\Sigma_n-\bm\Sigma_N)\bm t/2\}$. As $N\to\infty$, the partial sums converge in $L_2$, hence in distribution, and their characteristic functions converge to $\exp(-\bm t^\top\bm\Sigma_n\bm t/2)$, because $\bm\Sigma_N=\sum_{i>N}\bm w_i\bm w_i^\top$ is the tail of a convergent series and thus tends to $\bm 0$. Thus the limit $\sum_{i>n}\bm w_i\xi_i$ is distributed as $\mathcal N_p(\bm 0,\bm\Sigma_n)$.
Therefore, $\tilde{\bm\beta}\mid y_{1:n}\sim\mathcal{N}_p(\bm\beta_n,\bm\Sigma_n)$ for any $n\ge1$, and, taking $n=0$, $\tilde{\bm\beta}\sim\mathcal{N}_p(\bm\beta_0,\bm\Sigma_0)$.
The result for $\sigma_n^2=\sigma^2/a_n$ follows as a special case, since $\bm\Sigma_n=\sigma^2\bm V_n$ by Proposition~2.

\section{Additional simulations}

\subsection{Comparison between uncertainty quantification methods with Student-$t$ predictive distribution}

In this section, we consider a linear regression setting in which the response follows a location-scale Student-$t$ distribution with $\nu=5$ degrees of freedom and scale parameter $\phi=2$, both treated as known.
All other settings are as in Section~5.1 of the paper.
For each scenario, we generate 30 datasets.
In place of the exact posterior, which is no longer Gaussian, we use the Gaussian distribution $\mathcal{N}_p(\bm\beta_n,\bm\Sigma_n)$ with the exact posterior mean and covariance, where $\sigma^2=\phi^2\nu/(\nu-2)$ (\textsc{apx}); when $\sigma_n^2=\sigma^2a_n^{-1}$, this is the Gaussian approximation in Equation~(7) of the paper. Its $B$ draws provide the Monte Carlo reference for the errors.
The results, reported in Table~\ref{tab:sim_t}, are consistent with those for the Gaussian case.
The errors of \textsc{ps-1} and \textsc{ps-2} are comparable with the Monte Carlo errors of \textsc{apx}, and the widths and coverages of the credible intervals from predictive resampling essentially coincide with those of \textsc{apx}.
The coverage is nearly nominal when $\sigma_n^2=\sigma^2a_n^{-1}$; when $\sigma_n^2=\sigma^2$, the intervals are narrower and the coverage is slightly lower for $n=3p$.
Finally, \textsc{ps-2} is about 40 to 50 times faster than \textsc{ps-1}.

\begin{table}[t]
\centering
\caption{\small Comparison of uncertainty quantification from predictive resampling with Student-$t$ responses, with Algorithms~1 (\textsc{ps-1}) and~2 (\textsc{ps-2}) of the paper and with $B$ draws from the Gaussian distribution $\mathcal{N}_p(\bm\beta_n,\bm\Sigma_n)$ (\textsc{apx}): logarithm of the mean squared error of the posterior mean, $\log(\|\hat{\bm\beta}-\bm\beta_n\|^2/p)$, logarithm of the Frobenius norm of the lower-triangular part of $\sigma^{-2}(\hat{\bm\Sigma}-\bm\Sigma_n)$, average width and coverage of the 95\% intervals for the $p$ regression coefficients, and computation time in seconds. For \textsc{apx}, the errors are Monte Carlo errors and the intervals are the corresponding Gaussian ones. Values are averages over the simulated datasets, with standard deviations in parentheses.}
\vspace{0.2cm}
\label{tab:sim_t}
\footnotesize
\begin{tabular}{l l ccccc}
\hline
& & $\log(\|\hat{\bm\beta}-\bm\beta_n\|^2/p)$ & $\log\|\hat{\bm\Sigma}-\bm\Sigma_n\|_F$ & \textsc{width} & \textsc{cover} & \textsc{time}\\
\hline
\multicolumn{7}{l}{$n=3p$}\\
$\sigma_n^2=\sigma^2a_n^{-1}$ & \textsc{ps-1} & $-12.08$ & $-5.62$ & $1.006$ & $0.945$ & $6.81$ \\
& & $^{(0.22)}$ & $^{(0.04)}$ & $^{(0.012)}$ & $^{(0.036)}$ & $^{(0.46)}$ \\
 & \textsc{ps-2} & $-11.93$ & $-5.62$ & $1.007$ & $0.944$ & $0.14$ \\
& & $^{(0.30)}$ & $^{(0.03)}$ & $^{(0.012)}$ & $^{(0.033)}$ & $^{(0.01)}$ \\
 & \textsc{apx} & $-11.91$ & $-5.61$ & $1.010$ & $0.948$ & $-$ \\
& & $^{(0.26)}$ & $^{(0.03)}$ & $^{(0.012)}$ & $^{(0.033)}$ &  \\
\cmidrule{2-7}
$\sigma_n^2=\sigma^2$ & \textsc{ps-1} & $-12.23$ & $-5.74$ & $0.932$ & $0.927$ & $6.82$ \\
& & $^{(0.22)}$ & $^{(0.04)}$ & $^{(0.011)}$ & $^{(0.038)}$ & $^{(0.55)}$ \\
 & \textsc{ps-2} & $-12.08$ & $-5.77$ & $0.933$ & $0.928$ & $0.15$ \\
& & $^{(0.31)}$ & $^{(0.03)}$ & $^{(0.012)}$ & $^{(0.036)}$ & $^{(0.04)}$ \\
 & \textsc{apx} & $-12.06$ & $-5.76$ & $0.935$ & $0.927$ & $-$ \\
& & $^{(0.26)}$ & $^{(0.03)}$ & $^{(0.011)}$ & $^{(0.038)}$ &  \\
\hline
\multicolumn{7}{l}{$n=10p$}\\
$\sigma_n^2=\sigma^2a_n^{-1}$ & \textsc{ps-1} & $-13.49$ & $-7.12$ & $0.476$ & $0.948$ & $17.16$ \\
& & $^{(0.18)}$ & $^{(0.02)}$ & $^{(0.002)}$ & $^{(0.026)}$ & $^{(0.85)}$ \\
 & \textsc{ps-2} & $-13.52$ & $-7.12$ & $0.476$ & $0.946$ & $0.44$ \\
& & $^{(0.17)}$ & $^{(0.03)}$ & $^{(0.002)}$ & $^{(0.026)}$ & $^{(0.11)}$ \\
 & \textsc{apx} & $-13.43$ & $-7.11$ & $0.478$ & $0.949$ & $-$ \\
& & $^{(0.21)}$ & $^{(0.02)}$ & $^{(0.002)}$ & $^{(0.026)}$ &  \\
\cmidrule{2-7}
$\sigma_n^2=\sigma^2$ & \textsc{ps-1} & $-13.54$ & $-7.16$ & $0.464$ & $0.939$ & $17.66$ \\
& & $^{(0.17)}$ & $^{(0.02)}$ & $^{(0.002)}$ & $^{(0.029)}$ & $^{(0.73)}$ \\
 & \textsc{ps-2} & $-13.57$ & $-7.17$ & $0.465$ & $0.942$ & $0.44$ \\
& & $^{(0.17)}$ & $^{(0.03)}$ & $^{(0.002)}$ & $^{(0.027)}$ & $^{(0.01)}$ \\
 & \textsc{apx} & $-13.48$ & $-7.16$ & $0.466$ & $0.939$ & $-$ \\
& & $^{(0.21)}$ & $^{(0.02)}$ & $^{(0.002)}$ & $^{(0.028)}$ &  \\
\hline
\end{tabular}
\end{table}

\subsection{Comparison with frequentist additive models}
\label{apx:am}

In this section, we compare the proposed method with frequentist additive models (\textsc{am}), using the same settings as in the simulation study of Section~5.3 of the paper. We consider two versions of the additive model, both fitted with the \textsc{r} package \texttt{mgcv}: one with a Gaussian likelihood (\textsc{am-$n$}) and one with a location-scale Student-$t$ likelihood (\textsc{am-$t$}). 
We also include the Gaussian approximation to the martingale posterior given in Equation~(7) of the paper (\textsc{apx}). 
The methods are evaluated in terms of the mean squared error (\textsc{mse}) between the estimate of $\mathds{E}(Y \mid \bm x)$ and its true value on the test set, the average coverage and width of the 95\% intervals for $\mathds{E}(Y \mid \bm x)$ on the test set, and the total computation time. 
For the frequentist \textsc{am}, these intervals are confidence intervals rather than credible intervals.

The results are reported in Table~\ref{tab:sim_freq}. \textsc{am-$t$} yields the most accurate point estimates in both scenarios. 
The \textsc{rls} posterior mean has exactly the same form as the \textsc{am-$n$} point estimate. 
The two differ only in how the smoothing parameters are selected: \textsc{rls} maximizes the \textsc{loo-pll}, whereas \textsc{am-$n$} uses restricted maximum likelihood \citep{wood2011fast}.  
In both scenarios, all methods have coverage close to the nominal level, and the intervals obtained by predictive resampling are slightly narrower than those of the Gaussian approximation.
Finally, our method is faster than \texttt{mgcv}: the Gaussian approximation, which only requires the posterior mean, is more than ten times faster than both \textsc{am-$t$} and \textsc{am-$n$}, and \textsc{rls-$t$}, including predictive resampling, is more than four times faster than \textsc{am-$n$}.

\begin{table}[t]
\centering
\caption{\small Comparison between our proposed method with Student-$t$ predictive distribution, with predictive resampling (\textsc{rls-$t$}) and with the Gaussian approximation in Equation~(7) of the paper (\textsc{apx}), and frequentist additive models with Student-$t$ (\textsc{am-$t$}) and Gaussian (\textsc{am-$n$}) likelihoods, in terms of out-of-sample \textsc{mse}, average coverage and width of 95\% intervals for $\mathds{E}(Y\mid\bm x)$, and total computation time in seconds. Values are averages over the simulated datasets, with standard deviations in parentheses.}
\vspace{0.2cm}
\label{tab:sim_freq}
\footnotesize
\begin{tabular}{l cccc cc cccc}
\hline
& \multicolumn{4}{c}{$n=30p$} & & & \multicolumn{4}{c}{$n=100p$}\\
& {\textsc{rls-$t$}} & {\textsc{apx}} & {\textsc{am-$t$}} & {\textsc{am-$n$}} & & & {\textsc{rls-$t$}} & {\textsc{apx}} & {\textsc{am-$t$}} & {\textsc{am-$n$}}\\
\hline
\textsc{mse} & $0.438$ & $0.438$ & $\bm{0.383}$ & $0.434$ & & &
$0.119$ & $0.119$ & $\bm{0.104}$ & $0.120$ \\
& $^{(0.084)}$ & $^{(0.084)}$ & $^{(0.067)}$ & $^{(0.089)}$ & & &
$^{(0.014)}$ & $^{(0.014)}$ & $^{(0.012)}$ & $^{(0.015)}$   \\
\textsc{cover} & $0.935$ & $0.937$ & $0.944$ & $\bm{0.948}$ & & &
$0.948$ & $\bm{0.949}$ & $0.954$ & $0.959$ \\
& $^{(0.022)}$ & $^{(0.021)}$ & $^{(0.021)}$ & $^{(0.019)}$ & & &
$^{(0.014)}$ & $^{(0.014)}$ & $^{(0.015)}$ & $^{(0.013)}$   \\
\textsc{width ci} & $2.437$ & $2.448$ & $2.360$ & $2.547$ & & &
$1.331$ & $1.338$ & $1.282$ & $1.408$ \\
& $^{(0.128)}$ & $^{(0.129)}$ & $^{(0.101)}$ & $^{(0.124)}$ & & &
$^{(0.024)}$ & $^{(0.024)}$ & $^{(0.030)}$ & $^{(0.029)}$   \\
\textsc{time} & $0.72$ & $\bm{0.31}$ & $7.77$ & $3.37$ & & &
$2.00$ & $\bm{0.73}$ & $22.36$ & $9.14$ \\
& $^{(0.10)}$ & $^{(0.09)}$ & $^{(2.65)}$ & $^{(0.70)}$ & & &
$^{(0.51)}$ & $^{(0.08)}$ & $^{(7.63)}$ & $^{(1.76)}$   \\
\hline
\end{tabular}
\end{table}

\subsection{Additional details on effective sample size in simulations}

\begin{figure}[H]
\centering
\includegraphics[width=\textwidth]{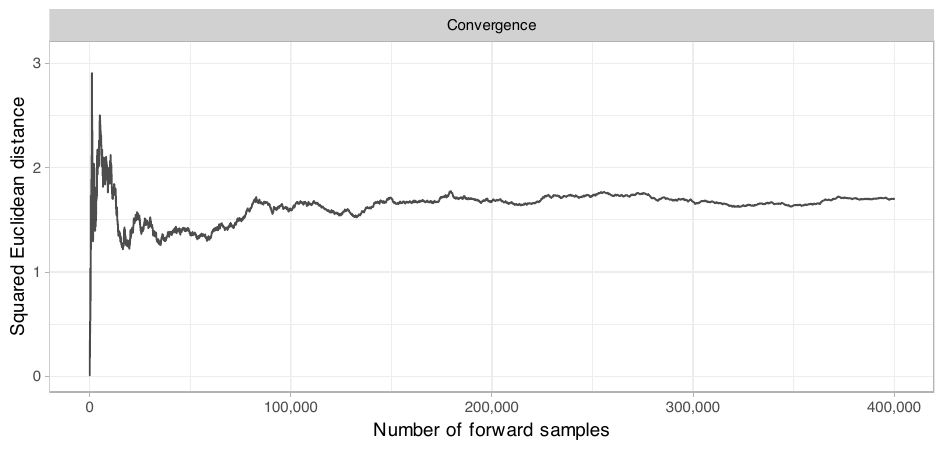}
\caption{\small One sample path of the squared Euclidean distance $\|\bm\beta_m-\bm\beta_n\|^2$ between $\bm\beta_m$, $m=n+1,\ldots,n+400{,}000$, and the initial~$\bm\beta_n$.}
\label{fig:additional_5}
\end{figure}

In this section, we detail how the lengths of the \textsc{mcmc} chains were chosen for the traditional Bayesian additive model and \textsc{bart} in the simulation study of Section~5.3 of the paper. We set the number of martingale posterior draws to $B=5{,}000$. Since Algorithm~2 of the paper runs separate forward chains of length $N$, the martingale posterior draws are independent by construction, given the shared future covariates, so their effective sample size (\textsc{ess}) equals $B$. The forward length $N$ can be chosen as described in Section~3.4 of the paper; Figure~\ref{fig:additional_5} shows a sample path of a forward chain for the analysis of Section~6 of the paper. For the competitors, we then chose the chain length and burn-in so that the average \textsc{ess} of the posterior draws of $f(\bm x)$ across the test set is close to $B$, when possible. For traditional Bayes, five chains of $2{,}000$ draws each, with the first $1{,}000$ discarded as burn-in, yield an \textsc{ess} close to $B$. \textsc{bart} mixes poorly, and its \textsc{ess} remains far below $B$ also with long chains; we run it for $55{,}000$ iterations, discarding the first $50{,}000$ as burn-in. The results are reported in Table~\ref{tab:sim_ess}.

\begin{table}[t]
\centering
\caption{\small Average effective sample size for the posterior draws of $f(\bm x)$ across the test set in the simulation study of Section~5.3 of the paper, for our proposed methods with Student-$t$ (\textsc{rls-$t$}) and Gaussian (\textsc{rls-$n$}) predictive distributions, the traditional Bayesian linear model with Student-$t$ likelihood (\textsc{blm}) and \textsc{bart}. Values are averages over the simulated datasets, with standard deviations in parentheses.}
\vspace{0.2cm}
\label{tab:sim_ess}
\footnotesize
\begin{tabular}{lc cccc}
\hline
& & {\textsc{rls-$t$}} & {\textsc{rls-$n$}} & {\textsc{blm}} & {\textsc{bart}}\\
\hline
$n=30p$ & & $5000$ & $5000$ & $\bm{5113.9}$ & $66.9$\\
& & $-$ & $-$ & $^{(51.6)}$ & $^{(15.1)}$\\
$n=100p$ & & $5000$ & $5000$ & $\bm{5286.4}$ & $56.9$\\
& & $-$ & $-$ & $^{(46.8)}$ & $^{(2.9)}$\\
\hline
\end{tabular}
\end{table}

\section{Pipeline for selecting the predictive distribution}

The validity of inference under the proposed method requires that the predictive distribution adequately reflect the unknown data-generating process. In this section, we describe a pipeline for selecting an appropriate predictive distribution. We illustrate it with the \textsc{nhanes} serum triglycerides dataset analyzed in Section~6 of the paper.
Although some choices at each step are specific to this application, the general ideas behind the pipeline apply more broadly.

\begin{figure}[t]
\centering
\includegraphics[width=\textwidth]{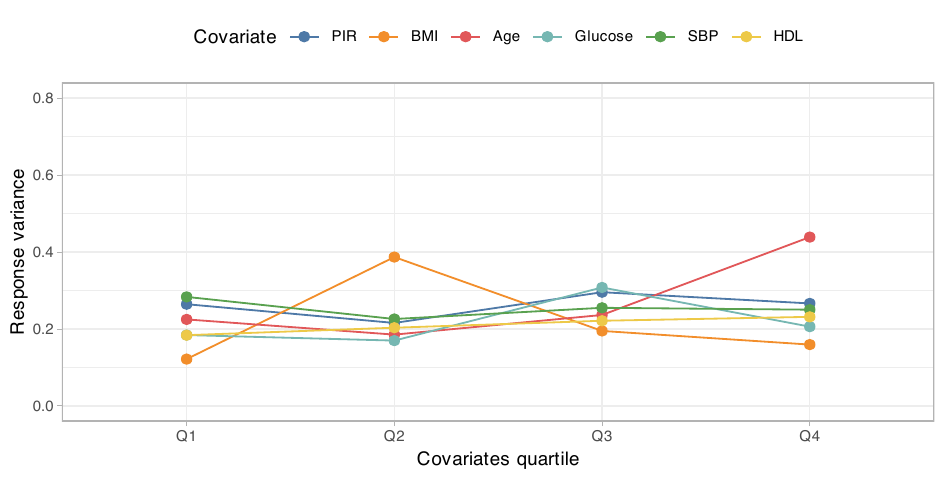}
\caption{\small Variance of log-triglycerides computed across quartiles of each continuous covariate.}
\label{fig:app_var}
\end{figure}

As an exploratory check of homoscedasticity, we divide the observations of the training set into quartiles of each continuous covariate, each containing between 16 and 26 observations, and compute the variance of log-triglycerides within each quartile; for the income ratio, which is top-coded at 5, the last group consists of the 26 top-coded observations. Ethnicity is not considered, since four of its six categories have fewer than ten observations. The results are reported in Figure~\ref{fig:app_var}. The variance of the response varies little across covariates and quartiles, remaining between 0.15 and 0.35 in all but three cases (0.12 and 0.39 for \textsc{bmi}, 0.44 for age), which are plausibly due to the small number of observations in each quartile. We therefore consider homoscedasticity a reasonable assumption for the response.

\begin{figure}[t]
\centering
\includegraphics[width=\textwidth]{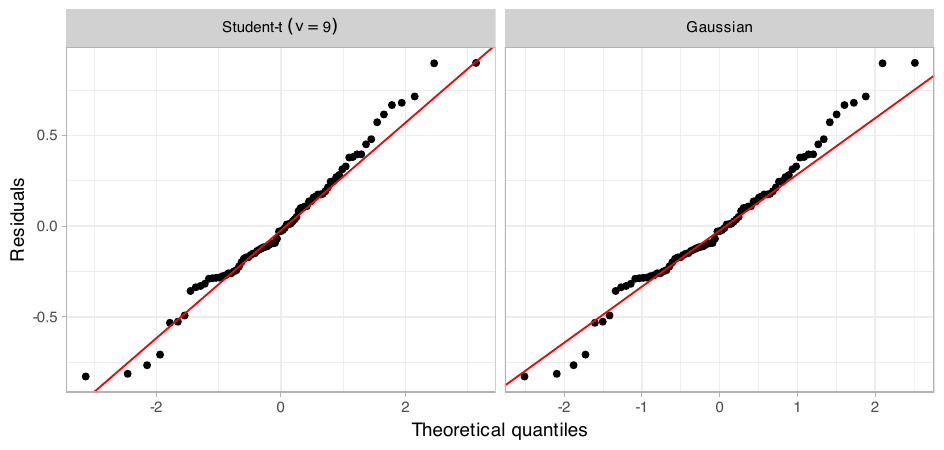}
\caption{\small Empirical quantiles of the residuals plotted against the theoretical ones of a Student-$t$ distribution with 9 degrees of freedom (left) and a Gaussian distribution (right).}
\label{fig:app_qq}
\end{figure}

Next, we assess whether a Gaussian predictive distribution is adequate or whether a heavier-tailed distribution is needed. To do so, we fit a frequentist additive model with a Gaussian likelihood using \texttt{mgcv} and compare the quantiles of the resulting residuals with the theoretical quantiles of candidate distributions. As shown in Figure~\ref{fig:app_qq}, for the \textsc{nhanes} serum triglycerides dataset, the residual quantiles are more aligned with those of a Student-$t$ distribution with 9 degrees of freedom than with those of a Gaussian distribution. We therefore adopt a Student-$t$ predictive distribution.

Finally, we consider whether a skew Student-$t$ distribution might be more appropriate than a symmetric one. We then maximize the \textsc{loo-pll} with $P_n$ taken to be a skew Student-$t$ distribution. The estimated degrees of freedom and scale are 13.65 and 0.36, respectively, very close to the estimates under the symmetric Student-$t$ reported in the main paper. The estimated skewness parameter is $-0.007$, thereby indicating that a symmetric Student-$t$ predictive distribution is sufficient.

\section{Analysis of NHANES serum triglycerides dataset: additional figures}
\label{apx:figures}

This section contains additional figures related to the analysis of \textsc{nhanes} serum triglycerides dataset in Section~6 of the paper.

\begin{figure}[H]
\centering
\includegraphics[width=\textwidth]{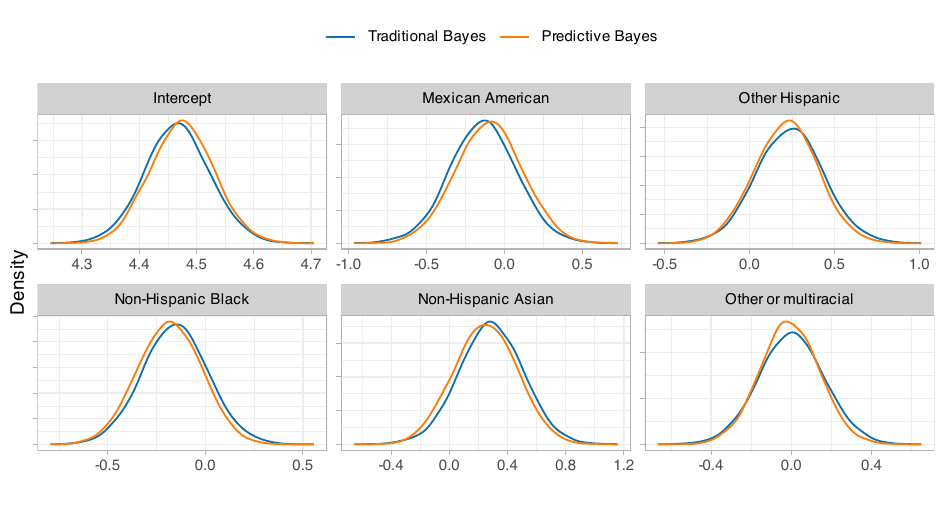}
\caption{\small Posterior distributions of the intercept and of the regression coefficients of the ethnicity indicators, with non-Hispanic white as the baseline, obtained with traditional and predictive Bayes.}
\label{fig:app_d}
\end{figure}

\begin{figure}[H]
\centering
\includegraphics[width=\textwidth]{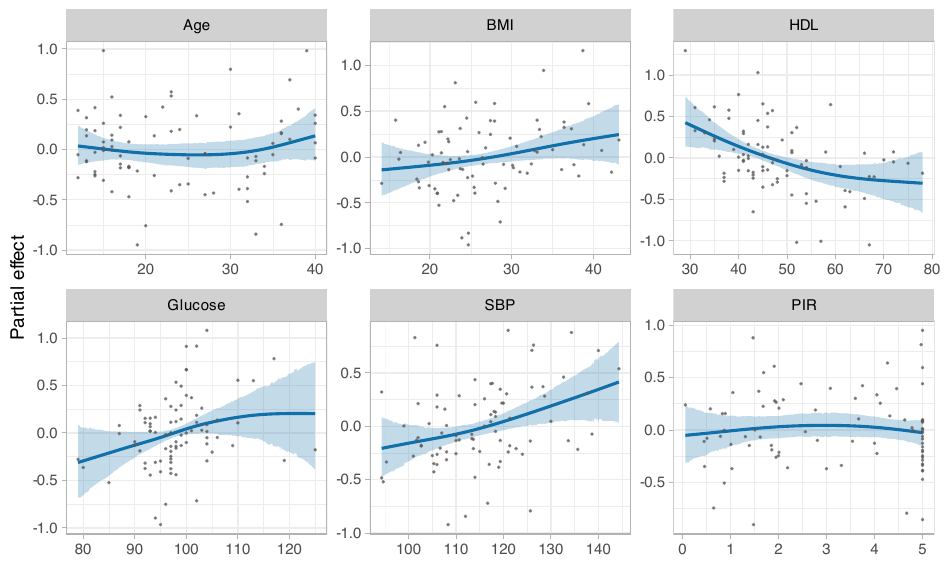}
\caption{\small Estimated covariate-specific effects with pointwise 95\% highest posterior density bands, obtained using the traditional Bayesian model, and partial residuals. The bands are truncated to the range of the partial residuals.}
\label{fig:app_blm}
\end{figure}

\begin{figure}[H]
\centering
\includegraphics[width=\textwidth]{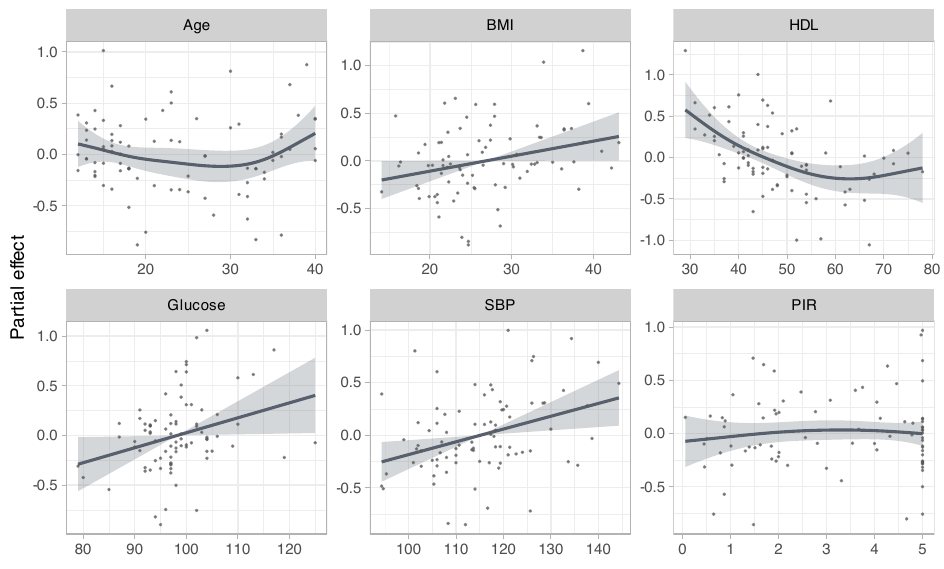}
\caption{\small Estimated covariate-specific effects with pointwise 95\% confidence bands, obtained using a frequentist additive model with Student-$t$ likelihood, and partial residuals. The bands are truncated to the range of the partial residuals.}
\label{fig:app_freq}
\end{figure}

\section{Computational details}

This section describes how the posterior mean, the \textsc{loo-pll} of Section~4.3 of the paper and its gradient, and the predictive resampling of Algorithm~2 are computed in our implementation. We present them for the additive models of Section~4.1 of the paper, with features $\bm z_i\in\mathds{R}^r$ and $\bm V_0=(\bm\Lambda+\varepsilon\bm I_r)^{-1}$; the linear model is the special case without smoothing parameters.

\subsection{Posterior mean and leave-one-out prequential log-likelihood}

The recursion of Equations~(3)--(4) of the paper is not needed on the observed data, since its output is available in closed form. Recalling Equation~(8) of the paper, $\bm V_n^{-1}=\bm A$ and $\bm\beta_n=\bm A^{-1}\bm Z^\top\bm y$, where
\begin{equation*}
\bm A:=\bm\Lambda+\varepsilon\bm I_r+\bm Z^\top\bm Z,
\end{equation*}
$\bm Z\in\mathds{R}^{n\times r}$ has rows $\bm z_1^\top,\ldots,\bm z_n^\top$ and $\bm y=(y_1,\ldots,y_n)^\top$. The matrix $\bm Z^\top\bm Z$ and the vector $\bm Z^\top\bm y$ do not depend on the hyperparameters, so they are computed once for each dataset. For given smoothing parameters, we compute the Cholesky factorization $\bm A=\bm R^\top\bm R$, with $\bm R$ upper triangular, and obtain $\bm\beta_n$ by two triangular solves. The leverages needed by the \textsc{loo-pll} are $h_i=\bm z_i^\top\bm V_n\bm z_i=\|\bm R^{-\top}\bm z_i\|^2$, so they follow from one further triangular solve with right-hand side $\bm Z^\top$. The leave-one-out predictions are then $\bm z_i^\top\bm\beta_{-i}=y_i-(y_i-\bm z_i^\top\bm\beta_n)/(1-h_i)$, as in Section~4.3 of the paper. Each evaluation of the \textsc{loo-pll} therefore costs $O(r^3+nr^2)$ operations.

\subsection{Gradient and maximization}

The hyperparameters are selected by maximizing the \textsc{loo-pll} over the logarithms of the smoothing parameters $\lambda_1,\ldots,\lambda_p$ and of the scale and shape parameters of $P_n$, with the quasi-Newton routine \texttt{nlminb} of \textsc{r}. For the additive models of Sections~5.3 and~6 of the paper, we supply the analytic gradient of the \textsc{loo-pll}, so that no finite differences are needed; the linear models of Section~5.2, whose only hyperparameters are the two or three parameters of $P_n$, use the finite differences of \texttt{nlminb}. The gradient follows from the chain rule, starting from the differentials of $\bm V_n$, $\bm\beta_n$ and $h_i$ with respect to $\log\lambda_j$. Since $\bm A$ depends on $\lambda_j$ only through the block $\lambda_j\bm\Lambda_j$ of $\bm\Lambda$, the rule $\mathrm{d}\bm A^{-1}=-\bm A^{-1}(\mathrm{d}\bm A)\bm A^{-1}$ for the differential of the inverse \citep{magnus2019matrix} gives the differentials of $\bm V_n=\bm A^{-1}$, of $\bm\beta_n=\bm A^{-1}\bm Z^\top\bm y$ and of $h_i=\bm z_i^\top\bm A^{-1}\bm z_i$. The derivatives with respect to the parameters of $P_n$ follow from those of its log-density; for the Student-$t$ distribution, the derivative with respect to the degrees of freedom involves the digamma function. All these quantities are obtained from the Cholesky factor $\bm R$ and from $\bm A^{-1}\bm Z^\top$, so the gradient costs the same order of operations as the \textsc{loo-pll} itself.

\subsection{Predictive resampling}

In Algorithm~2 of the paper, the covariates $\bm X_{n+1},\ldots,\bm X_N$ are sampled from the empirical distribution of the observed features, and the weights $\bm w_i$ and matrices $\bm V_i$ are computed once, in $O(Nr^2)$ operations. Each of the $B$ draws then requires sampling $N-n$ innovations from $P_\xi$ and computing a weighted sum, in $O(Nr)$ operations, and the draws are computed in parallel.

Stopping the series at $N$ omits the variance $\bm\Sigma_N$ of the remainder $\tilde{\bm\beta}-\bm\beta_N=\sum_{i>N}\bm w_i\,\xi_i$, given the covariates. With $\sigma_n^2=\sigma^2/a_n$, Section~3.2 of the paper shows that $\bm\Sigma_N=\sigma^2\bm V_N$. When $\bm V_0^{-1}$ is negligible, $\bm V_N$ is roughly $n/N$ times $\bm V_n$, as noted in Section~3.4 of the paper, so $n/N$ gives the order of the omitted fraction of the posterior variance.

The last step of Algorithms~1 and~2 adds to each draw an independent draw from $\mathcal{N}_p(\bm 0,\sigma^2\bm V_N)$, in place of the remainder. Two facts justify it. First, the remainder is approximately Gaussian: it is a weighted sum of infinitely many independent innovations, each with a small weight, and Theorem~2 of the paper, applied with $N$ in place of $n$, shows that it is asymptotically Gaussian with covariance $\bm\Sigma_N$. The non-Gaussian features of the martingale posterior, such as heavy tails, come mainly from the first terms of the series, which have the largest weights and are all simulated. Second, the covariance $\sigma^2\bm V_N$ is exact with $\sigma_n^2=\sigma^2/a_n$, as recalled above. With a constant $\sigma_n=\sigma$, Lemma~\ref{lem:sandwich} with $N$ in place of $n$ and $s_N^-=s_N^+=\sigma^2$ gives
\begin{equation*}
\sigma^2(1-\epsilon_N)\bm V_N\preceq\bm\Sigma_N\preceq\sigma^2\bm V_N,
\qquad \epsilon_N=\sup_{i>N}\bm x_i^\top\bm V_{i-1}\bm x_i.
\end{equation*}
The quantity $\epsilon_N$ can be bounded by observed quantities. For $i>N$, Equation~(4) of the paper gives $\bm V_{i-1}^{-1}=\bm V_N^{-1}+\sum_{k=N+1}^{i-1}\bm x_k\bm x_k^\top\succeq\bm V_N^{-1}$, and inversion reverses $\preceq$, so $\bm V_{i-1}\preceq\bm V_N$. Hence $\epsilon_N\le\max_{\bm x}\bm x^\top\bm V_N\bm x$, where the maximum is over the support of the distribution of the future covariates, that is, over the observed features when they are resampled. Therefore, adding a draw from $\mathcal{N}_p(\bm 0,\sigma^2\bm V_N)$ overstates the covariance of the remainder by at most $\epsilon_N\sigma^2\bm V_N$, where $\epsilon_N$ is of order $1/N$ and computable.

In all the examples of the paper, we use $N=10n$ with the correction. We checked this choice against a reference with a much larger $N$, for which the omitted part of the posterior variance is negligible: $N=1{,}000n$ in the simulation study of Section~5.3 of the paper, on five datasets per scenario, $N=n+100{,}000$ in the two examples of Section~5.2, and $N=n+300{,}000$ in the application of Section~6. In all cases, the 95\% credible intervals are close to the reference ones: in Section~5.3, the average width and coverage of the intervals for $f(\bm x)$ coincide with the reference ones to three decimal places; in Section~5.2, the widths of the intervals for the coefficients are between $0.99$ and $1.01$ times the reference ones and their endpoints differ by at most $3.5\%$ of the width, also in the first example, whose Student-$t$ $P_n$ has $\hat\nu=2.60$ and $n=15$; in Section~6, the widths of the intervals for $\mathds{E}(Y\mid\bm x)$ at the test points are $1.004$ times the reference ones on average, and their endpoints differ by at most $2\%$ of the width. Without the correction, the omitted part of the posterior variance is of order $n/N=10\%$, and the intervals are narrower, by about $5\%$ in Sections~5.1 and~5.2 and $6\%$ to $8\%$ in Section~5.3.

\clearpage
\bibliography{biblio}

\end{document}